\documentclass[12pt,a4paper]{article}

\usepackage{amsmath,xcolor}
\usepackage{amssymb}
\usepackage{epsfig}
\usepackage{subfigure}

\newcommand{\re}{{\mathop{\rm Re}}\,}
\newcommand{\im}{{\mathop{\rm Im}}\,}
\newcommand{\kam}{\textsc{kam}}
\newcommand{\D}{{\rm d}}
\newcommand{\E}{{\rm e}}
\newcommand{\I}{{\rm i}}

\newcommand{\R}{\mathbb{R}}

\newcommand{\Z}{\mathbb{Z}}

\newcommand{\C}{\mathbb{C}}

\newcommand{\sS}{\mathbb{S}}

\newcommand{\cA}{\mathcal A}
\newcommand{\cB}{\mathcal B}
\newcommand{\cC}{\mathcal C}
\newcommand{\cD}{\mathcal D}

\newcommand{\cK}{\mathcal K}
\newcommand{\cO}{\mathcal O}
\newcommand{\cP}{\mathcal P}
\newcommand{\cQ}{\mathcal Q}
\newcommand{\cU}{\mathcal U}
\newcommand{\cV}{\mathcal V}

\newtheorem{theorem}{Theorem}[section]

\newtheorem{rmrk}[theorem]{Remark}
\newtheorem{xmpl}[theorem]{Example}

\newcommand{\qed}{\nolinebreak\hfill {$\Box$} \par\medbreak}
\newcommand{\finis}{\nolinebreak\hfill {$\bigtriangleup$} \par\medbreak}
\newenvironment{remark}{\begin{rmrk} \begin{rm}}{\end{rm} \end{rmrk}}

\newcommand{\opin}[2]{\mathopen] #1, #2 \mathclose[}

\newcommand{\bs}{\begin{subequations}}    % for lazy typers
\newcommand{\es}{\end{subequations}}
\newcommand{\be}{\begin{equation}}
\newcommand{\ee}{\end{equation}}
\newcommand{\bd}{\begin{displaymath}}
\newcommand{\ed}{\end{displaymath}}
\newcommand{\ba}{\begin{eqnarray}}
\newcommand{\ea}{\end{eqnarray}}
\newcommand{\bas}{\begin{eqnarray*}}
\newcommand{\eas}{\end{eqnarray*}}

\newcommand{\blue}{\textcolor{blue}}
\newcommand{\red}{\textcolor{red}}

\begin{document}

\title{\protect\Large On the detuned $2{:}4$ resonance}

\author{{\protect\normalsize Heinz Han{\ss}mann} \protect\\[-1mm]
   {\protect\footnotesize\protect\it Mathematisch Instituut,
           Universiteit Utrecht} \protect\\[-2mm]
   {\protect\footnotesize\protect\it Postbus 80010,
           3508~TA Utrecht, The Netherlands} \protect\\
   {\protect\normalsize Antonella Marchesiello} \protect\\[-1mm]
   {\protect\footnotesize\protect\it Faculty of Information Technology,
           Czech Technical University in Prague} \protect\\[-2mm]
   {\protect\footnotesize\protect\it Th\'{a}kurova 9,
           16000 Prague, Czech Republic} \protect\\
   {\protect\normalsize Giuseppe Pucacco} \protect\\[-1mm]
   {\protect\footnotesize\protect\it Dipartimento di Fisica,
           Universit\`a Tor Vergata Roma} \protect\\[-2mm]
   {\protect\footnotesize\protect\it Via della Ricerca Scientifica 1,
           00133 Roma, Italy}}

%\date{\protect\normalsize  {18 February 2020}}
% date written, not date \latexed, please update 

\maketitle

\begin{abstract}
\noindent
We consider families of Hamiltonian systems in two degrees of
freedom with an equilibrium in $1{:}2$~resonance.
Under detuning, this ``Fermi resonance'' typically leads to normal
modes losing their stability through period-doubling bifurcations.
For cubic potentials this concerns the short axial orbits and
in galactic dynamics the resulting stable periodic orbits are
called ``banana'' orbits.
Galactic potentials are symmetric with respect to the co-ordinate
planes whence the potential --- and the normal form --- both have
no cubic terms.
This $\Z_2 \times \Z_2$--symmetry turns the $1{:}2$~resonance into
a higher order resonance and one therefore also speaks of the
$2{:}4$~resonance.
In this paper we study the $2{:}4$~resonance in its own right, not
restricted to natural Hamiltonian systems where $H = T + V$
would consist of kinetic and (positional) potential energy.
The short axial orbit then turns out to be dynamically stable
everywhere except at a simultaneous bifurcation of banana and
``anti-banana'' orbits, while it is now the long axial orbit that
loses and regains stability through two successive period-doubling
bifurcations.

\end{abstract}

\section{Introduction}
\label{sec:introduction}

Symmetries play a fundamental role in the mathematical modeling of
physical systems.
Either exact or approximate, they produce extra conservation laws or
constrain the structure of relevant equations indicating the way to
solve the problem at hand~\cite{kozlov}.
A particularly striking example is provided by Hamiltonian systems
close to resonance around an elliptic equilibrium.
The structure of the normal form is largely determined by discrete
symmetries affecting the degree of the lowest order resonant
terms~\cite{tv}.
Consider in two degrees of freedom the lowest-order genuine
$1{:}2$~resonance~\cite{CDHS}: its prototype is the Fermi
resonance and a simple mechanical example is the
{\em spring-pendulum}~\cite{Broer1998}.
When enforcing approximate reflectional symmetry with respect to both
degrees of freedom a higher-order normal form becomes necessary.
Indeed, the cubic resonant terms are removed from the normal form and
the first non-vanishing resonant terms are of $6$th order --- squaring
cubic terms yields invariance under reflections.
We follow~\cite{Contopoulos2004} and denote the resulting problem as
$2{:}4$~resonance.
However, it shares several features of the lowest-order case and can
be investigated with analogous techniques.

A classical example is that of the motion of a star in an elliptical
galaxy whose gravitational potential possesses mirror reflection with
respect to each symmetry plane \cite{Verhulst1979, MP2013b}.
When the flattening is small, motion in the core is well approximated
by a perturbed symmetric $1{:}1$~oscillator \cite{zm,PM14}.
But when the flattening is high, the dynamics can be closer to the
symmetric $1{:}2$~resonance \cite{MS,MP2013a}.
Axial orbits of arbitrary amplitude exist and may suffer instability
at some threshold.
At such a threshold a periodic orbit in
{\em general position}~\cite{SVM07}
bifurcates off from the axial orbit together with a symmetric
counterpart forming a mirror-symmetric pair.
This has interesting consequences for the structure of the system. 

\begin{remark}
For the banana orbits it is straightforward to ``see'' the two
mirror-symmetric members of the pair.
The two trajectories of the anti-banana (figure-eight) pair are
instead simply going in opposite direction on the same orbit in
configuration space.
\end{remark}

\noindent
Let us consider a family of $\Z_2 \times \Z_2$--symmetric
Hamiltonian systems in two degrees of freedom close to an elliptic
equilibrium, which is equivariant with respect to the
reflectional symmetries 
\bs 
\label{symmetries}
\begin{align}
   \varrho_1 \; : & \;\; (x_1,x_2,y_1,y_2) \;\;
   \mapsto \;\; (-x_1,x_2,-y_1,y_2)
\label{symmetry-1}\\
   \varrho_2 \; : & \;\; (x_1,x_2,y_1,y_2) \;\;
   \mapsto \;\; (x_1,-x_2,y_1,-y_2)
\label{symmetry-2}
\end{align}
\es
where $(x,y)$ denote the canonical co-ordinates.
Assuming the Hamiltonian to be an analytic function in a
neighbourhood of the equilibrium, its series expansion about
the equilibrium point can be written as
\be
\label{Hamiltonian series}
H(x, y; \delta) \;\; = \;\;
\sum_{j=0}^{\infty} H_{2j}(x, y; \delta)
\ee
where $H_{2j}$ are homogeneous polynomials of degree~$2(j+1)$
in the co-ordinates~$(x, y)$; we discuss the dependence on
the parameter $\delta \in \R$ below.
Note that in force of the reflectional symmetries~\eqref{symmetries},
odd degree terms are not present in the expansion.
The quadratic part
\be
\label{H0}
H_0(x, y; \delta) \;\; = \;\;
\frac{\omega_1}{2}(x_1^2 + y_1^2) \; + \;
\frac{\omega_2}{2}(x_2^2 + y_2^2)
\ee
of~\eqref{Hamiltonian series} describes two oscillators with
frequencies $\omega_j = \omega_j(\delta) \in \R$, $j = 1, 2$ coupled
by the nonlinear terms in~\eqref{Hamiltonian series} which we
consider as a perturbation of~\eqref{H0}.
The dynamics of the linear Hamiltonian system defined by~\eqref{H0}
is readily analysed.
The $(x_1, y_1)$--plane and the $(x_2, y_2)$--plane both consist of
periodic orbits and the rest of the phase space is foliated by
invariant $2$--tori.
Our aim is to understand what happens under addition of higher order
terms.
Persistence of invariant $2$--tori is addressed by \kam~Theory.
The linear approximation $H_0$ of~$H$ has a constant frequency
mapping and thus fails to satisfy the Kolmogorov
condition~\eqref{Kolmogorov}.
To obtain an integrable approximation of~$H$ that does satisfy the
Kolmogorov condition we compute a truncated normal form~$K$
with respect to~$H_0$, see~\cite{BP} and references therein.
If there are no resonances 
\begin{displaymath}
   k_1 \omega_1 \; + \; k_2 \omega_2 \;\; = \;\; 0
   \enspace , \quad  0 \neq k \in \Z^2
\end{displaymath}
of order $|k| := |k_1| + |k_2| \leq n$,
then the normal form of order~$n$ depends on~$(x, y)$ only as a
function of the invariants
\be
\label{invariants}
   \tau_1 \; = \; \frac{x_1^2 + y_1^2}{2}
   \quad \mbox{and} \quad
   \tau_2 \; = \; \frac{x_2^2 + y_2^2}{2}
   \enspace .
\ee
Such a {\em Birkhoff normal form} 
\begin{displaymath}
   K \;\; = \;\; \omega_1 \tau_1 \; + \; \omega_2 \tau_2
   \; + \; \frac{\omega_{11}}{2} \tau_1^2 \; + \;
   \omega_{12} \tau_1 \tau_2 \; + \;
   \frac{\omega_{22}}{2} \tau_2^2 \; + \; \ldots
\end{displaymath}
generically satisfies the Kolmogorov condition
\begin{equation}
\label{Kolmogorov}
   \det (\omega_{ij})_{ij} \;\; = \;\;
   \omega_{11} \omega_{22} \; - \; \omega_{12}^2
   \;\; \neq \;\; 0
\end{equation}
and/or the iso-energetic non-degeneracy condition
\begin{equation}
\label{isoenergetic}
   2 \omega_{12} \omega_1 \omega_2 \, - \, \omega_{11} \omega_2^2
   \; - \; \omega_{22} \omega_1^2 \;\; \neq \;\; 0 \enspace .
\end{equation}
Therefore, our analysis of the unperturbed system essentially
remains valid for~\eqref{Hamiltonian series}.
The non-degeneracy conditions \eqref{Kolmogorov}
and~\eqref{isoenergetic} require the computation of the Birkhoff
normal form of order $n \geq 4$, hence these considerations do
not apply to resonances up to order~$4$.
Next to the $1{:}1$ and $1{:}{-}1$~resonances this excludes
the $1{:}{\pm}2$ and $1{:}{\pm}3$~resonances.
Correspondingly, in~\cite{Sanders1978} all other resonances are
called higher order resonances.

In two degrees of freedom all normal forms are integrable,
but the normal form truncated at order~$4$ of a resonance
of order $|k| \leq 4$ generically contains extra
``resonant terms'' of order~$|k|$.
The resulting dynamics depend thus on the lower-order
resonance at hand.
Also for higher-order resonances a reliable approximation of
the dynamics of~\eqref{Hamiltonian series} might require the
normalization to be performed at least up to the order at
which the first resonant term appears~\cite{Contopoulos2004}.

\subsection{ Approach to the resonance}
\label{sec:approach}

Here we consider the problem of determining the phase space
structure of the perturbation of a $1{:}2$~resonant oscillator
invariant under the symmetries~\eqref{symmetries}.
To catch the main features of the orbital structure, we make our
parameter~$\delta$ a {\em detuning} parameter~\cite{Verhulst1979}
by assuming 
\be
\label{detuning}
\omega_1 \;\; = \;\;
\left( \frac{1}{2} \; + \; \delta \right) \omega_2
\ee
and proceed as if the unperturbed harmonic part were in exact
$1{:}2$~resonance, thus including the detuning inside the
perturbation.
In this way we turn~\eqref{H0} into
\bs
\label{pert12oscillator}
\begin{align}
   & H_0 \;\; = \;\; 
   \omega_2 \left( \frac{\tau_1}{2} \; + \; \tau_2 \right)
\label{12oscillator}
\intertext{while the perturbation becomes}
\label{perturbation}
   & + \; \delta \omega_2 \tau_1 \; + \; 
   \sum_{j=1}^{\infty} H_{2j}(x, y; \delta)
\end{align}
\es
where the dependence of $H_{2j}$, $j \geq 1$ on~$\delta$ may be
arbitrary, e.g.\ polynomial; for definiteness we assume that the
parameter $\delta \in \R$ only appears in the detuning with all
$H_{2j} = H_{2j}(x, y)$ independent of~$\delta$ (and discuss
below in how far this captures the behaviour of general
$1$--parameter families).

We aim at a general understanding of the bifurcation sequences of
periodic orbits in general position from the normal modes,
parametrised by the ``energy''~$E$, the detuning parameter~$\delta$
and the independent coefficients characterising the nonlinear
perturbation.
This problem was already studied in the case of
``natural Hamiltonians''~\cite{MP2013a}, i.e.\ in case the potential
depends only on the ``spatial'' variables~$x$, and therefore
$H_{2j}=H_{2j}(x)$ for $j \geq 1$.
Here we consider the more general system~\eqref{pert12oscillator}.
We follow a different, geometric approach that allows not only
to reproduce the results of~\cite{MP2013a}, but also to extend
these and, under certain assumptions, to deduce the generic
behavior of~\eqref{pert12oscillator}.
The results obtained are summarized in
theorems \ref{thm: instability normal modes}
and~\ref{thm: bifurcation sequences}.
Actually the value~$E$ of the Hamiltonian~$H$ does not always
correspond to the energy of the system now, but colloquially
we shall still call $E$ the (generalized) energy. 

As remarked above, in presence of symmetries the minimal truncation
order necessary to include at least one resonant term in the
normal form depends not only on the order~$|k|$ of the resonance,
but also on the symmetries at hand.
For the reflectional symmetries~\eqref{symmetries} the minimal
truncation order increases to $2|k|$,
see~\cite{Contopoulos2004, han07, SVM07}.
Thus, in this point of view, the symmetric $1{:}2$~resonance
behaves as a higher order resonance, and as said we shall speak
of $2{:}4$~resonance.

\subsection{ What is new}
\label{sec:whatisnew}

The approach we take to study the $2{:}4$~resonance has become
rather standard, compare
with~\cite{meer85, CFH99, ILPPSY06, MP14, EHM18} and references
therein.
Normalizing about the periodic flow of the resonant oscillator
introduces an extra continuous symmetry, cf.~\cite{BP, SVM07},
while preserving already existing symmetries of the system.
Studying the normal form dynamics in their own right allows to
reduce to one degree of freedom, cf.~\cite{CB1997, efs04}.
We follow the treatment of resonant normal modes in the
$3D$~H\'enon--Heiles family in~\cite{HvdM02} and first consider
the insufficient $4$th order normal form before turning to the
$6$th order normal form necessary for the fine structure, see
also~\cite{tv}.
Aspects of the dynamics that are persistent under addition of
higher order normalized terms have a chance to persist also when
``perturbing back'' to the original system (of which normal forms
of increasingly high order form an increasingly close approximation).

 The normal form turns out to have an
$\sS^1 \times \Z_2$--symmetry,
where the second factor is inherited from the second factor of the
original $\Z_2 \times \Z_2$--symmetry and the first factor~$\sS^1$
is an improvement upon the original~$\Z_2$ due to normalization.
Reducing this symmetry by means of invariants allows to get a global
picture of the (reduced) dynamics, see
Figs.~\ref{Fig: singular equilibria}, \ref{fig3} and~\ref{Fig: X2ab}
below.
The cuspidal form of the singularity corresponding to the family of
short axial orbits explains why here bifurcations of banana and
anti-banana orbits now happen simultaneously.
This perspective also allows to decide at once that going to any
higher order than~$6$ in the normalization process does not again
lead to qualitative changes, but only to quantitative ones.

We introduce the (truncated) normal form for the
system~\eqref{pert12oscillator} in section~\ref{sec:reduction}
and reduce the dynamics to one degree of freedom. 
We do this in two steps, first reducing the $\sS^1$--symmetry and
then the remaining $\Z_2$--symmetry.
Then by a geometric approach we study the equilibria of the reduced
system and describe the possible bifurcation sequences in sections
\ref{sec:first order approx} and~\ref{sec:second order approx}. 
In section~\ref{sec:first order approx} we restrict to the normal
form of order~$4$ while the improvements due to the normal form of
order~$6$ are presented in section~\ref{sec:second order approx}.
The results so obtained are used in section~\ref{sec:bifurcations}
to deduce the dynamics of the original system.
Section~\ref{sec:example} demonstrates our results for a specific
class of examples.
Some final comments and conclusions follow in
section~\ref{sec:conclusions}.

\section{Reduction}
\label{sec:reduction}

Let us zoom in on the neighbourhood of the equilibrium at the
origin by introducing a perturbing parameter $\varepsilon > 0$,
scaling co-ordinates as
\bs
\label{scaling}
\begin{align}
\label{epsilon scaling}
(x,y) & \;\; \mapsto \;\; (\varepsilon x, \varepsilon y)
\intertext{and also the detuning~\eqref{detuning} as}
\label{detuning scaling}
\delta & \;\; \mapsto \;\; \varepsilon^2 \delta \enspace,
\intertext{so that it can be treated as a second order term in the
perturbation.
Scaling furthermore time as}
\label{time scaling}
t & \;\; \mapsto \;\; \frac{\varepsilon^2\omega_2}{2} \, t
\end{align}
\es
no $\varepsilon$ remains in the unperturbed resonant
oscillator~\eqref{12oscillator} while we get $\omega_2 = 2$ for the
frequencies in the Hamiltonian~\eqref{pert12oscillator}, thereby
turning~\eqref{pert12oscillator} into
\be
\label{Hamiltonian epsilon}
H \;\; = \;\; \tau_1 \; + \; 2 \tau_2
\; + \; 2 \varepsilon^2 \delta \tau_1 \; + \;
\sum_{j=1}^{\infty}\varepsilon^{2j}H_{2j}
\enspace .
\ee
The system defined by~\eqref{Hamiltonian epsilon} is in general not
integrable, even after truncation of the convergent series.
The flow~$\varphi^{H_0}_t$ of the unperturbed
system~\eqref{12oscillator} yields the $\sS^1$--action
$\varphi^{H_0}$ on~$\R^4 \cong \C ^2$ given by
\be
\label{oscillator symmetry}
\begin{array}{cccc}
   \varphi^{H_0} \; : & \sS^1\times\C^2 & \longrightarrow & \C^2 \\
   & (\ell, (z_1, z_2) ) & \mapsto &
   (\E^{- \I \ell} z_1, \E^{- 2 \I \ell } z_2)
\end{array}
\ee
where
\begin{displaymath}
   z_j \;\; = \;\; x_j \; + \; \I y_j
   \enspace , \quad  j = 1,2 \enspace .
\end{displaymath}
The perturbed Hamiltonian~\eqref{Hamiltonian epsilon} is in general
not invariant under this action, however we can normalize~$H$ so
that the resulting normal form does have the oscillator
symmetry~\eqref{oscillator symmetry}.
A set of generators of the Poisson algebra of
$\varphi^{H_0}$--invariant functions is given by 
\begin{displaymath}
   \tau_1 \;\; = \;\; \frac{z_1 \bar{z}_1}{2}
   \enspace , \quad
   \tau_2 \;\; = \;\; \frac{z_2 \bar{z}_2}{2}
\end{displaymath}
introduced in~\eqref{invariants} together with
\be
\label{tau generators}
   \sigma_1 \; = \; \frac{\re z_1^2 \bar{z}_2}{2}
   \enspace , \quad
   \sigma_2 \; = \; \frac{\im z_1^2 \bar{z}_2}{2}
\ee
and it is constrained by $\tau_1 \geq 0$, $\tau_2 \geq 0$ and the
syzygy
\be
\label{tau constraints}
R(\tau, \sigma) \;\; := \;\; 2 \tau_1^2 \tau_2 \; - \;
(\sigma_1^2 + \sigma_2^2) \;\; = \;\; 0 \enspace .
\ee
See~\cite{CB1997, CDHS, han07} for more details.
The normalization allows us to reduce the dynamics to one degree
of freedom as the Poisson bracket on~$\R^4$ induced
by \eqref{invariants} and~\eqref{tau generators} has two
Casimir elements, namely $R$ and $H_0 = \tau_1 + 2 \tau_2$.
For a fixed value $\eta \geq 0$ of~$H_0$ we can eliminate
$\tau_2 = \frac{1}{2}(\eta - \tau_1)$.
The dynamics are constrained to the reduced phase space
\be
\label{reduced phase tau}
\cV^{\eta} \;\; = \;\; \left\{ \,
(\tau_1, \sigma_1, \sigma_2) \in \R^3 \; : \;
R^{\eta}(\tau_1, \sigma_1, \sigma_2) = 0,
\; 0 \leq \tau_1 \leq \eta \, \right\}
\ee
with Poisson structure
\begin{displaymath}
   \{ f, g \} \;\; = \;\;
   \langle \nabla f \times \nabla g \mid \nabla R^{\eta} \rangle
   \enspace ,
\end{displaymath}
where
\begin{displaymath}
   R^{\eta}(\tau_1, \sigma_1, \sigma_2) \;\; = \;\;
   (\eta - \tau_1) \tau_1^2 \; - \; (\sigma_1^2 + \sigma_2^2)
   \enspace .
\end{displaymath}
The normal form for the
$2{:}4$~resonance~\eqref{Hamiltonian epsilon},
truncated at order~$6$ in the original variables~$(x, y)$,
has the general structure
\be
\label{general normal form}
K(\tau, \sigma; \delta) \;\; = \;\; K_0(\tau) \; + \;
\varepsilon^2 K_2(\tau; \delta) \; + \; \varepsilon^4 \left[
\mu \frac{\sigma_1^2 - \sigma_2^2}{2} \, + \,
\nu \sigma_1 \sigma_2 \, + \, K_4(\tau; \delta) \right]
\ee
with
\bas
 K_0 & = & H_0 \;\; = \;\; \tau_1 \; + \; 2 \tau_2 \;\; = \;\; \eta \\
 K_2 & = & 2 \delta \tau_1 \; + \; \alpha_1 \tau_1^2
 \, + \, \alpha_2 \tau_2^2 \, + \, \alpha_3 \tau_1 \tau_2 \\
 K_4 & = & \rho_1 \delta \tau_1^2 \, + \, \rho_2 \delta \tau_2^2
 \, + \, \rho_3 \delta \tau_1 \tau_2
 \; + \; \alpha_4 \tau_1^2 \tau_2 \, + \, \alpha_5 \tau_1 \tau_2^2
 \, + \, \alpha_6 \tau_1^3 \, + \, \alpha_7 \tau_2^3
\enspace . 
\eas
The coefficients $\mu, \nu$, $\rho_j$, $j = 1, 2 ,3$ and $\alpha_i$,
$i = 1, \ldots, 7$ depend on the coefficients of the polynomial
terms~$H_{2j}$ in the original Hamiltonian~\eqref{Hamiltonian epsilon}.
To keep our analysis as general as possible, here and in the following
we prefer to work with the normal form~\eqref{general normal form}
with the most generic coefficients.
Afterwards, in section~\ref{sec:example} we give an application
to an explicit class of systems.
We assume at least one of the coefficients $\mu$ and~$\nu$ to be
non-vanishing, otherwise the first order at which the
normal form yields stabilized dynamics would be higher.

\begin{remark}
The $\delta$--dependent terms with coefficients $\rho_j$ in~$K_4$
are an artefact of the normalization procedure and if we decide to
normalize to higher order also the resulting $K_6, K_8, \ldots$
are going to depend on the detuning~$\delta$.
In case the $H_{2j}$ in~\eqref{perturbation} do depend on~$\delta$,
we develop these dependencies into series and adjust the passage
from the $H_{2j}$ in~\eqref{pert12oscillator} to the
$H_{2j}$ in~\eqref{Hamiltonian epsilon} according to the
scaling~\eqref{detuning scaling}.
While this does affect the quantitative values of  the~$\rho_j$,
once these changed values are computed there are no
further adjustments to be made and in particular the qualitative
statements in the sequel remain unchanged.
\end{remark}

\noindent
The $\Z_2 \times \Z_2$--symmetry of~\eqref{Hamiltonian epsilon}
generated by~\eqref{symmetries} is inherited by the normal
form~\eqref{general normal form}.
In fact, for $\ell = \pi$ the $\sS^1$--action~\eqref{oscillator symmetry}
yields the reflectional symmetry~\eqref{symmetry-1}; correspondingly,
none of the invariants in \eqref{invariants} and~\eqref{tau generators}
changes under~\eqref{symmetry-1}.
The remaining symmetry~\eqref{symmetry-2} becomes
\be
\label{sigma symmetries}
(\tau, \sigma) \;\; \mapsto \;\; (\tau, -\sigma)
\ee
whence the normal form~\eqref{general normal form} depends on
$\sigma_1, \sigma_2$ only via $\frac{1}{2}(\sigma_1^2 - \sigma_2^2)$
and~$\sigma_1 \sigma_2$.
We perform a further reduction to explicitly divide out this symmetry,
by introducing variables~\cite{HS}
\be
\label{Z coordinates}
  \begin{array}{rcl}
    u & := & \tau_1 \\
    v & := & \frac{1}{2}(\sigma_1^2 - \sigma_2^2) \\
    w & := & \sigma_1 \sigma_2  \enspace .
  \end{array}
\ee
Note that, since the reduced phase space is a surface of revolution,
by rotation we can always eliminate one of the two variables~$v, w$
from the Hamiltonian (recall that we do not consider the
case $\mu = \nu = 0$ here).
For definiteness we assume from now on $\mu > 0$ and $\nu = 0$.

\begin{remark}
\label{remark222}
If the system is reversible, then $\nu = 0$ from the start, but
$\mu$ might be negative and in applications it is not always helpful
to actually perform a $\pi$--rotation to achieve $\mu > 0$.
Therefore we sometimes also comment on the case $\mu < 0$.
For the same reason we do not simply scale to $\mu = 1$.
\end{remark}

\noindent
The normal form~\eqref{general normal form} then becomes, after
neglecting constant terms and scaling one more time
by~$\varepsilon^2$,
\be
\label{reduced Hamiltonian w}
K^{\eta}(u, v, w; \delta) \;\; = \;\;
(2 \delta + \alpha \eta) u \, + \, \lambda u^2 \; + \;
\varepsilon^2 \left[\mu v \, + \, K^{\eta}_4(u; \delta) \right]
\ee
where 
\be
\label{newK4}
K_4^{\eta}(u; \delta) \;\; = \;\; \beta_1 \delta u^2 \; + \;
\beta_2 \delta \eta u \; + \; \gamma_1 u^3 \; + \;
\gamma_2 \eta u^2 \; + \; \gamma_3 \eta^2 u
\ee
and
\bs
\label{new coefficients} 
\begin{align}
&\lambda = \alpha_1 + \frac{\alpha_2}{4} - \frac{\alpha_3}{2},
\;\; \alpha = \frac{\alpha_3}{2} - \frac{\alpha_2 }{2},
\;\; \beta_1 = \rho_1 + \frac{\rho_2}{4} - \frac{\rho_3}{2},
\;\; \beta_2 = \frac{\rho_3}{2} - \frac{\rho_2}{2},
\label{new coefficients lab}\\
&\gamma_1 =
- \frac{\alpha_4}{2} + \frac{\alpha_5}{4} + \alpha_6 - \frac{\alpha_7}{8},
\;\; \gamma_2 =
\frac{\alpha_4}{2} - \frac{\alpha_5 }{2} + \frac{3 \alpha_7}{8},
\;\; \gamma_3 = \frac{\alpha_5}{4} - \frac{3 \alpha_7}{8}.
\label{new coefficients gamma} 
\end{align}
\es
To prevent that for any $\eta \geq 0$ there is the detuning
$\delta = - \frac{1}{2} \alpha \eta$ for which
$K^{\eta} = \cO(\varepsilon^2)$ we make the genericity assumption
$\lambda \neq 0$ on the coefficients of the $4$th order terms
$H_2(x, y)$ in~\eqref{Hamiltonian series}/\eqref{Hamiltonian epsilon}.

\begin{remark}
\label{remark22}
We follow a perturbative approach, introducing $\varepsilon$ as a
small perturbing parameter and looking for the bifurcation curves
in the $(\delta,\eta)$--plane as power series in~$\varepsilon$.
The results obtained are reliable only for small values
of~$\varepsilon$, i.e.\ if the original system is not too far from
the equilibrium at the origin and if the resonance
ratio~\eqref{detuning} is not too far from the $2{:}4$~resonance
 --- the detuning being scaled as in~\eqref{detuning scaling}.
\end{remark}

\noindent
The (twice) reduced phase space
\be
\label{reduced phase w}
\cP^{\eta} \;\; = \;\; \left\{ \, (u, v, w) \in \R^3 \; : \;
S^{\eta}(u, v, w) = 0, \; 0 \leq u \leq \eta \, \right\}
\ee
has the Poisson structure
\begin{displaymath}
   \{ f, g \} \;\; = \;\;
   \langle \nabla f \times \nabla g \mid \nabla S^{\eta} \rangle
   \enspace ,
\end{displaymath}
where
\begin{displaymath}
   S^{\eta}(u, v, w) \;\; = \;\; \frac{(\eta - u)^2}{2} u^4
   \; - \; 2 (v^2 + w^2)  \enspace .
\end{displaymath}
Correspondingly, the equations of motion take the form
\begin{displaymath}
   \frac{\D}{\D t} \left(
   \begin{array}{c}
      u \\ v \\ w
   \end{array}
   \right) \;\; = \;\;
   \nabla K^{\eta} \; \times \; \nabla S^{\eta}
\end{displaymath}
on $\cP^{\eta} \subseteq \R^3$ whence the singular points
$\cQ_1 := (0, 0, 0)$ and $\cQ_2 := (\eta, 0, 0)$ are always
equilibria for the reduced system.
The corresponding isotropy subgroups of the $\sS^1 \times \Z_2$--action
combining \eqref{oscillator symmetry} with~\eqref{sigma symmetries}
are given in table \ref{table:isotropy}.

\begin{table}
\begin{center}
\begin{tabular}{|c|c|c|c|c|}
  \hline
  $\C^2$ & $\cP^{\eta}$ & isotropy subgroup & dynamics & period \\
   \hline
  $\frac{1}{2} |z_1|^2 = \eta \neq 0$, $z_2 = 0 = \tau_2$ & $\cQ_2$ &
  $\Z_2$ & first normal mode & long \\
  $\tau_1 = 0 = z_1$, $|z_2|^2 = \eta \neq 0$ & $\cQ_1$ &
  $\{ 0, \pi \} \times \Z_2$ & second normal mode & short \\
  \hline
  \end{tabular}
  \end{center}
  \caption{\small Isotropy subgroups of the
    $\sS^1$--action~\eqref{oscillator symmetry}, $H_0 = \eta \neq 0$,
    combined with the $\Z_2$--action~\eqref{sigma symmetries}.
    Note that in case $\eta \to 0$ the phase space~$\cP^{\eta}$ shrinks
    to the equilibrium at the origin, with corresponding isotropy subgroup
    $\sS^1 \times \Z_2$.
  } \label{table:isotropy}
\end{table}

\begin{remark}
Note that $H_0$ is an integral of motion for the reduced
system~\eqref{general normal form} and not for the original
system~\eqref{Hamiltonian epsilon}, which is in general not
integrable, the Hamiltonian being its only integral of motion.
In~\cite{Broer1998} the value $\eta$ of~$H_0$
is also referred to as distinguished parameter.
In section~\ref{sec:bifurcations} we shall describe the bifurcations
of the original system in terms of the detuning~$\delta$ and of the
(generalized) energy~$E$.
\end{remark}

\noindent
We aim at understanding the dynamics on the reduced phase
space~$\cP^{\eta}$.
In particular, we look for the critical curves in the
$(\delta,\eta)$--plane corresponding to the bifurcations, together
with the possible bifurcation sequences (which would then depend also
on the coefficients of the normal form, actually not on all of them,
as we shall see).
Then, we use the information obtained from the normal form to deduce
the dynamics of the original system.
We start by investigating the equilibria of the system defined
by~\eqref{reduced Hamiltonian w}.

\section{First order approximation}
\label{sec:first order approx}

In this section we treat the $2{:}4$~resonance as a higher order
resonance.
As there are no cubic resonance terms, this means that we work with
an approximating Birkhoff normal form of order~$4$ in the original
variables~$(x, y)$, i.e.\ we first look at its first order
approximation, obtained by neglecting the second order term
in~$\varepsilon$ of~\eqref{reduced Hamiltonian w} and study
the dynamics defined by
\be
\label{reducedHamiltonianwepszero}
K_{\delta}^{\eta}(u, v, w) \;\; = \;\;
(2 \delta + \alpha \eta) u \; + \; \lambda u^2
\ee
on~\eqref{reduced phase w} where $\delta, \eta$ are parameters
with $\eta$ distinguished with respect to~$\delta$ and $\lambda, \alpha$
are non-vanishing constant coefficients.
Note that this puts the conditions $\alpha_3 \neq \alpha_2$ and
$\alpha_3 \neq 2 \alpha_1 + \frac{1}{2} \alpha_2$ on the~$\alpha_i$
in~\eqref{general normal form}, but there are no conditions on
$\beta$ and~$\gamma$ in~\eqref{newK4}; recall that $\mu > 0$
in~\eqref{reduced Hamiltonian w}.
Later on we furthermore require $\alpha + 2 \lambda \neq 0$,
a genericity assumption that puts the additional constraint
$\alpha_3 \neq 4 \alpha_1$ on the~$\alpha_i$ in~\eqref{general normal form}.

\subsection{The equilibria on the reduced phase space}
\label{sec:equilibria}

\begin{figure}[htb]
\begin{center}
\begin{picture}(214,132)
   \put(0,-10){\includegraphics[height=5cm,keepaspectratio]{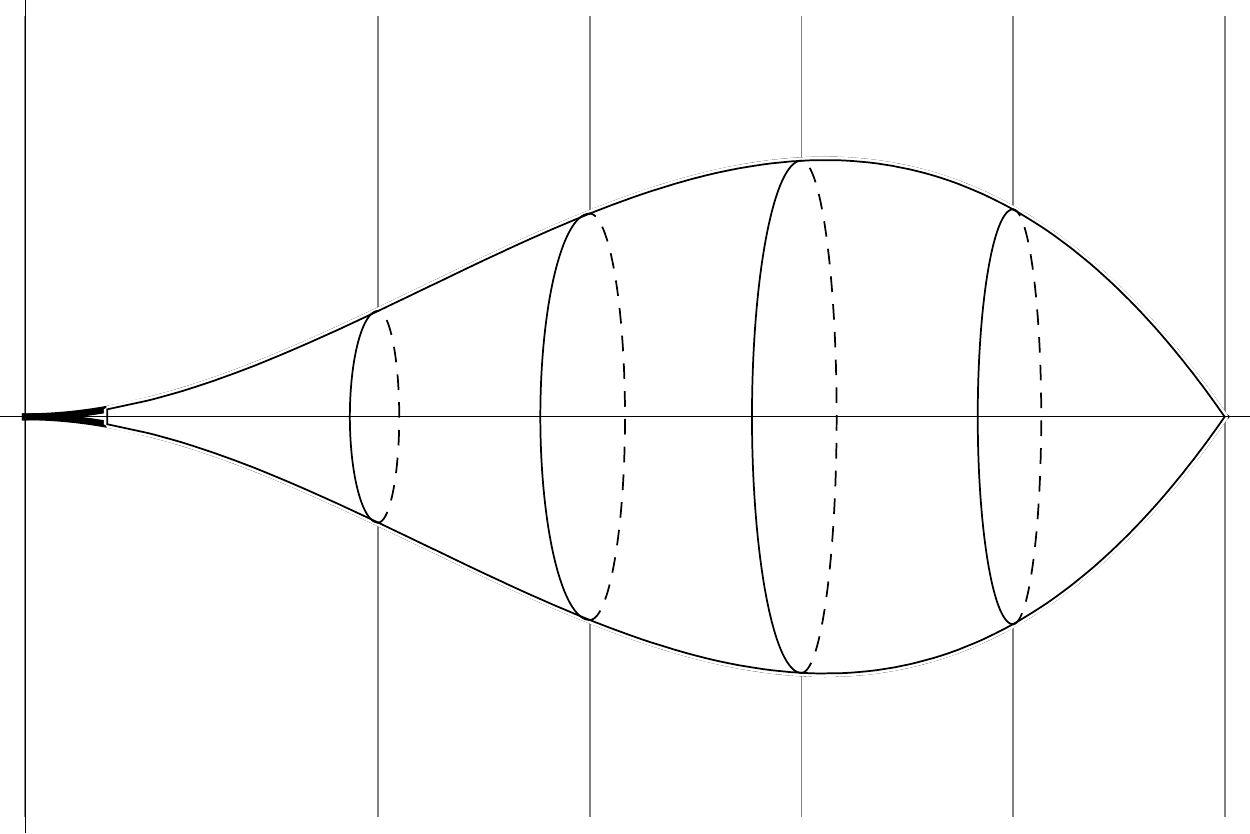}}
   \put(212,54){\footnotesize $u$}
   \put(9,128){\footnotesize $v$}
\end{picture}
\end{center}
\caption{\small
Possible intersections between the level sets~\eqref{level sets h}
and the reduced phase space.
Such intersections correspond to stable singular equilibria
or to periodic orbits.
Here we depict only level sets of~\eqref{level sets h}
that do not degenerate into a double plane (for those the circle
consists of equilibra).
\label{Fig: singular equilibria}}
\end{figure}

The reduced phase space~\eqref{reduced phase w} has a cuspidal
singularity at $\cQ_1 = (0, 0, 0)$ and a conical singularity
at $\cQ_2 = (\eta, 0, 0)$ and these are always equilibria (see
Fig.~\ref{Fig: singular equilibria}).
The intersections of~$\cP^{\eta}$ with the level sets 
\be
\label{level sets h}
\cK_{\delta}^{\eta}(h) \;\; := \;\; \left\{ \, (u, v, w) \in \R^3 \; : \;
K_{\delta}^{\eta}(u, v, w) = h \, \right\}
\ee
for
\begin{displaymath}
   h \;\; = \;\; K_{\delta}^{\eta}(\cQ_i)
   \enspace , \quad i = 1, 2 
\end{displaymath}
consist of the isolated~$\cQ_i$, whence both equilibria are stable.

\begin{remark}
The origin~$\cQ_1$ reconstructs to the family of short axial orbits
as predicted by Lyapunov's Centre Theorem~\cite{SVM07} and is singular
already on~$\cV^{\eta}$.
For the $1{:}2$~resonance the point
$(\tau_1, \sigma_1, \sigma_2) = (\eta, 0, 0)$ --- which corresponds to
the family of long axial orbits and gets reduced to~$\cQ_2$ --- is not
a singular point of~$\cV^{\eta}$ and correspondingly Lyapunov's Centre
Theorem does not apply here.
The extra $\Z_2$--symmetry turns the $1{:}2$~resonance into a
$2{:}4$~resonance whence also the family of long axial orbits becomes
a normal mode, see again~\cite{SVM07}.
The relation between normal modes and singular points of the reduced
phase space extends to $n$~degrees of freedom and we refer
to~\cite{MPY18} for more information.
\end{remark}

\noindent
The remaining (non-empty) intersections
\be
\label{level sets h_0}
\cP^{\eta} \; \cap \; \cK_{\delta}^{\eta}(h) \;\; \subseteq \;\; \R^3
\enspace , \quad
h \; = \; (2 \delta + \alpha \eta) u \, + \, \lambda u^2
\enspace , \quad
0 < u < \eta
\ee
yield ``great circles'' on the surface of revolution~$\cP^{\eta}$ as
the level sets $\cK_{\delta}^{\eta}(h)$ consist of two vertical planes
perpendicular to the $u$--axis (recall that we assumed $\lambda \neq 0$).
From the equations of motion
\bas
\dot{u} & = & 0 \\
\dot{v} & = & 4 w \frac{\partial K_{\delta}^{\eta}}{\partial u} \\
\dot{w} & = & - 4 v \frac{\partial K_{\delta}^{\eta}}{\partial u}
\eas
we infer that these great circles are periodic orbits, except when
$\cK_{\delta}^{\eta}(h)$ is a double plane where the circle consists
of equilibria.
Since
\bd
\frac{\partial K_{\delta}^{\eta}}{\partial u} \;\; = \;\;
2 \delta \; + \; \alpha \eta \; + \; 2 \lambda u
\ed
the corresponding double root is given by
\be
\label{zero order tangency}
u \; = \; u_0 \; := \;
- \frac{2 \delta + \alpha \eta}{2 \lambda}
\ee
and it gives a circle on the reduced phase space only if
\be
\label{ex condition tangency}
0 \; < \; u_0 \; < \; \eta
\enspace .
\ee
This restricts the parameter values to
\bs
\label{higherorderparameters}
\begin{align}
   \cD_{\lambda}^{\alpha} & \;\; := \;\; \left\{ \,
   (\delta, \eta) \; : \; - \lambda \eta - \frac{\alpha \eta}{2}
   < \delta < - \frac{\alpha \eta}{2} \, \right\}
   \qquad \mbox{if $\lambda > 0$}
\label{higherorderparameterspositive}\\
   \cD_{\lambda}^{\alpha} & \;\; := \;\; \left\{ \,
   (\delta, \eta) \; : \; - \frac{\alpha \eta}{2} < \delta <
   | \lambda | \eta - \frac{\alpha \eta}{2} \, \right\}
   \qquad \mbox{if $\lambda < 0$}
\label{higherorderparametersnegative}
\end{align}
\es
and outside of the closure $\overline{\cD_{\lambda}^{\alpha}}$
of~$\cD_{\lambda}^{\alpha}$ the dynamics is indeed what is
expected~\cite{Sanders1978} from a higher order resonance: the
phase flow consists of a family of periodic orbits extending between
the two singular equilibria, which therefore must be stable (see
Fig.~\ref{Fig: singular equilibria}).
Higher order terms in~$\varepsilon$ clearly change the shape of the
intersections~\eqref{level sets h_0}.
However, for $\varepsilon$ small enough the dynamics qualitatively
stays the same.
In two degrees of freedom the singular equilibria reconstruct
to the two normal modes and the periodic orbits reconstruct to a single
family of invariant $2$--tori satisfying the Kolmogorov and the
iso-energetic non-degeneracy condition.
We have recovered the description of the dynamics given in the
introduction, which indeed is valid for all higher order resonances. 

The dynamics become more intricate (and interesting) if the equation
defining~\eqref{level sets h_0} has two coinciding roots, i.e.\
where $(\delta, \eta) \in \cD_{\lambda}^{\alpha}$.
In this case, second order terms in the reduced
Hamiltonian~\eqref{reduced Hamiltonian w} are not negligible and they
are needed to describe the phase portrait of the system. 
We defer this full treatment of~\eqref{reduced Hamiltonian w},
with $\varepsilon > 0$, to section~\ref{sec:second order approx}
below.
Note that for $\lambda = 0$, which we excluded, the reduced phase
space~$\cP^{\eta}$ consists of equilibria
of~\eqref{reducedHamiltonianwepszero} when
$2 \delta + \alpha \eta = 0$ and then {\em all} aspects of the
dynamics of~\eqref{reduced Hamiltonian w} are determined by the
higher order terms.

\subsection{The bifurcation diagram}
\label{sec:bifurcation diagram}

\begin{figure}[htb]
\begin{center}
\begin{picture}(140,135)
   \put(0,-10){\includegraphics[height=5cm,keepaspectratio]{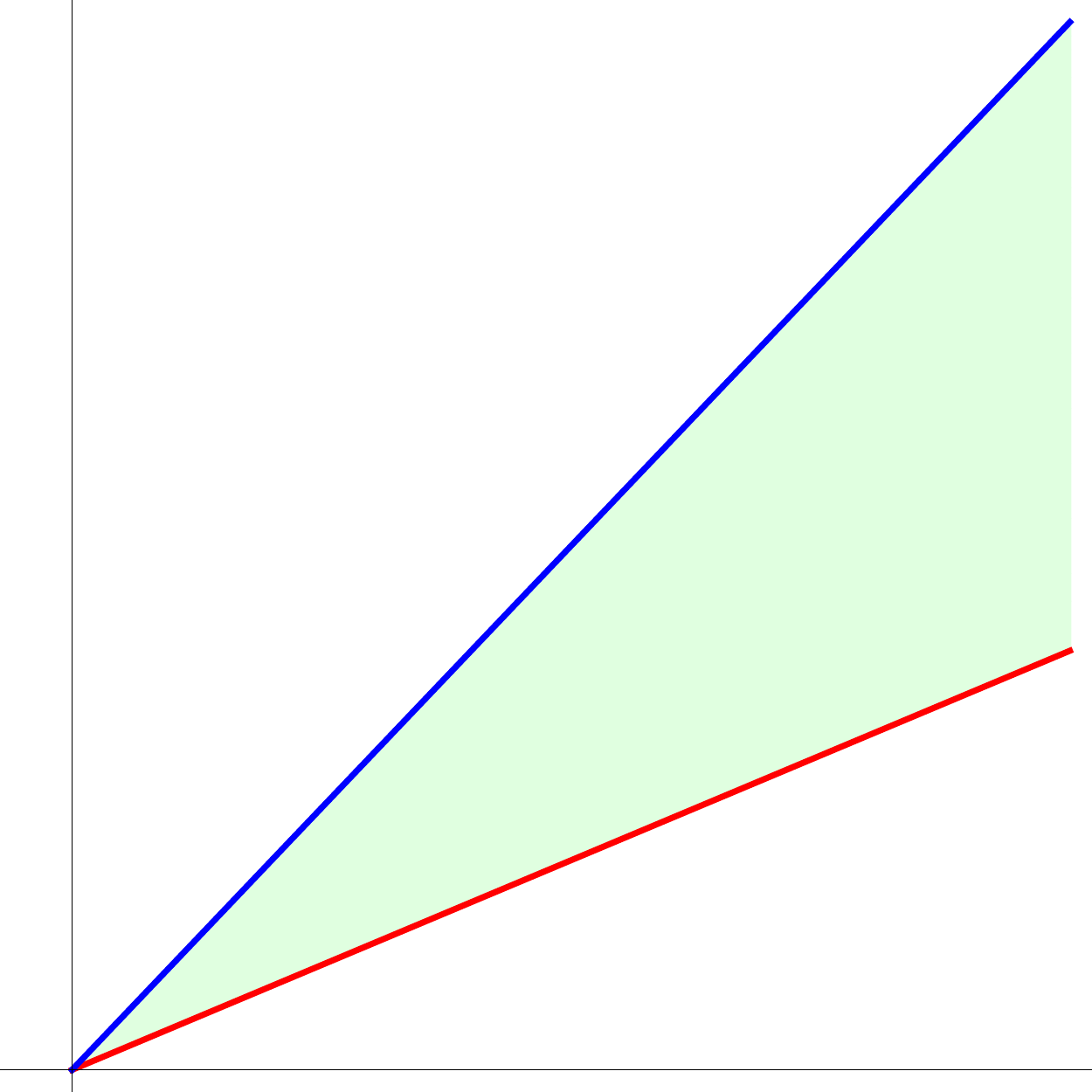}}
   \put(136,-5){\footnotesize $\delta$}
   \put(13,126){\footnotesize $\eta$}
\end{picture}
\end{center}
\centering
\caption{\small
Bifurcation diagram for $\varepsilon = 0$, depicted with
parameter values $\alpha = -0.5$ and $\lambda = 0.15$.
Along the red line the equilibrium~$\cQ_1$ is degenerate.
The green sector represents the set $\cD_{\lambda}^{\alpha}$
defined in~\eqref{higherorderparameters} --- for
$(\delta, \eta) \in \cD_{\lambda}^{\alpha}$
the dynamics has one great circle of regular equilibria next
to the singular equilibria $\cQ_1$ and~$\cQ_2$.
The equilibrium~$\cQ_2$ is degenerate along the blue line.
In Fig.~\ref{fig: daDiagram} this corresponds to the
case~\blue{V}-\red{III} (below-middle).
}\label{fig2}
\end{figure}

The reduced dynamics on~\eqref{reduced phase w} is governed by the
parameters $\delta$ and~$\eta$, the latter being distinguished with
respect to the former, while the coefficients
$(\lambda, \alpha) \in \R^2$ obtained from the Birkhoff normal form
via~\eqref{new coefficients lab} determine the shape
of~$\cD_{\lambda}^{\alpha}$ and thus where bifurcations take place.
Indeed, the double planes pass through the singular point
$\cQ_1 = (0, 0, 0) \in \cP^{\eta}$ when $2 \delta + \alpha \eta = 0$
and through $\cQ_2 = (\eta, 0, 0) \in \cP^{\eta}$ when
$2 \delta + (\alpha + 2 \lambda) \eta = 0$.
This yields the bifurcation diagram in Fig.~\ref{fig2}, with
structurally stable dynamics on~$\cP^{\eta}$ for
$(\delta, \eta) \notin \overline{\cD_{\lambda}^{\alpha}}$
and a great circle of equilibria for
$(\delta, \eta) \in \cD_{\lambda}^{\alpha}$.

Note that the red boundary cannot be vertical since $\alpha \neq 0$
and the requirement $\alpha + 2 \lambda \neq 0$, i.e.\
$\alpha_3 \neq 4 \alpha_1$, ensures that the blue boundary cannot be
vertical as well.
This is an additional genericity assumption, ensuring that
next to the value
\begin{equation}
\label{etaatcusp}
   \eta_0 \; = \; - \frac{2 \delta}{\alpha}
   \quad \mbox{at} \quad \cQ_1
\end{equation}
also the value
\begin{equation}
\label{etaatcone}
   \eta_0 \; = \; - \frac{2 \delta}{\alpha + 2 \lambda}
   \quad \mbox{at} \quad \cQ_2
\end{equation}
is finite.
This requires $\delta \alpha \leq 0$ and $\delta (\alpha + 2 \lambda) \leq 0$,
respectively, since $\eta_0$ cannot be negative.

The boundary~$\partial \cD_{\lambda}^{\alpha}$ marks the transition
from the regime where the $2{:}4$~resonance behaves as a higher
order resonance to the regime where the inclusion of a $6$th order
resonance term to dissolve the continuum of equilibria becomes
crucial.
A bifurcation sequence along a straight line passing
through~$\cD_{\lambda}^{\alpha}$ consists of the structurally stable
flow developing a degenerate singular equilibrium
at~$\partial \cD_{\lambda}^{\alpha}$, the resulting great circle
of equilibria moving through the reduced phase space~$\cP^{\eta}$
to the other singular equilibrium which then becomes degenerate and
leading back to the structurally stable flow, but now with the
direction of the periodic orbits reversed.

\section{Second order approximation}
\label{sec:second order approx}

By \kam~Theory, most of the invariant $2$--tori reconstructed from
the {family of} great circles extending between the two
singular equilibria $\cQ_1$ and~$\cQ_2$ persist the perturbation
from~\eqref{reducedHamiltonianwepszero} to the original
system~\eqref{Hamiltonian epsilon}, while it is generic for resonant
tori to break up and not persist the perturbation.
The great circles that consist of equilibria already break up
under the integrable perturbation
from~\eqref{reducedHamiltonianwepszero}
to~\eqref{reduced Hamiltonian w}, subject to the genericity
conditions $\mu > 0$ and $\lambda \neq 0$ (recall that we
furthermore assume $\alpha \neq 0$ and $\alpha + 2 \lambda \neq 0$).
We therefore aim at understanding the dynamics around the
degenerate case
\be
\label{critical h_0}
h \;\; = \;\; h_0 \;\; := \;\;
- \frac{(2 \delta + \alpha \eta )^2}{4 \lambda }
\ee
when the equation in~\eqref{level sets h_0} has two coinciding
roots \eqref{zero order tangency}
satisfying~\eqref{ex condition tangency}.
What happens when we look at the normal form up to second order
terms in the perturbation, i.e.\
at~\eqref{reduced Hamiltonian w} with $\varepsilon > 0$, is that
single vertical planes~$\cK_{\delta}^{\eta}(h)$, $h$ away from~$h_0$,
get replaced by almost vertical surfaces that still lead to
intersections with the reduced phase space~$\cP^{\eta}$ that are
periodic orbits, while near the double vertical
planes~$\cK_{\delta}^{\eta}(h_0)$ these level sets become almost
parabolic cylinders, touching $\cP^{\eta}$ at elliptic equilibria
where the rest of the level set lies outside of~$\cP^{\eta}$ and at
hyperbolic equilibria where part of the level set lies inside of
$\cP^{\eta} \subseteq \R^3$.
For energy levels between these equilibria, the parabolic cylinders
intersect $\cP^{\eta}$ in periodic orbits circling around such an
elliptic equilibrium.

From the elliptic and hyperbolic equilibria the so-called banana
and anti-banana orbits are reconstructed.
In astronomical systems the stable orbits are usually called bananas
and the unstable ones anti-bananas.
Here we consider more general systems, and prefer to follow a
different nomenclatura, by calling anti-bananas the figure eight
orbits that correspond to tangencies on the upper part of~$\cP^{\eta}$.
We call banana orbits the orbits corresponding to tangencies on the
lower part of~$\cP^{\eta}$, independent of whether they are stable or
not; compare with~\cite{MS}.

\subsection{Singular equilibria and their stability}
\label{sec:singularequilibriaandtheirstability}

Let us start by investigating the stability of the singular equilibria.
In particular, we want to find the critical values of~$\eta$ (if any)
that correspond to a stability/instability transition of the singular
equilibria.
Indeed, while the mechanism how this happens is more transparent
when varying~$\delta$, the parameter~$\eta$ is distinguished with
respect to the detuning~$\delta$ and this point of view allows to look
at bifurcations when solely changing the initial conditions.
Such instability transitions produce new (regular) equilibria for
the reduced system, bifurcating off from the singular equilibria.
If the corresponding critical values of~$\eta$ are not too high,
this reflects in the bifurcation of periodic orbits from the normal
modes in the original system.
We shall discuss this point in sections \ref{sec:bifurcation sequences}
and~\ref{sec:bifurcations}.
Since we are now looking at the system near $h = h_0$, we
consider the level sets
\bd
\cK_{\delta, \varepsilon}^{\eta, h_0}(k) \;\; := \;\;
\left\{ \, (u, v, w) \in \R^3 \; : \;
K^{\eta}(u, v, w; \delta) = h_0 + \varepsilon^2 k
\, \right\}
\ed
which give a family of third order curves when intersecting
with the $(u, v)$--plane, with equation
\be
\label{generic curve}
v(u) \;\; = \;\;
\frac{1}{\mu} \left[ k \, - \,
\frac{\lambda}{\varepsilon^2}(u - u_0)^2
\, - \, K_4^{\eta}(u; \delta) \right]
\enspace ,
\ee
where $u_0$ was obtained in~\eqref{zero order tangency} in the first
order approximation.
The $\varepsilon^2$ in the denominator lets the parabolic part of the
curve~\eqref{generic curve} dominate over the cubic part~$K_4^{\eta}$.

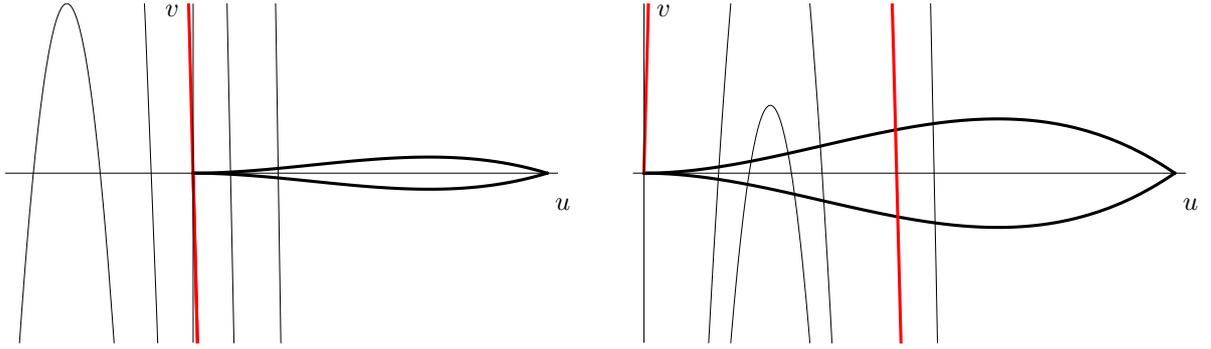
\begin{figure}[htb]
\begin{center}
\begin{picture}(435,120)
   \put(0,-10){\includegraphics[height=4.5cm,keepaspectratio]{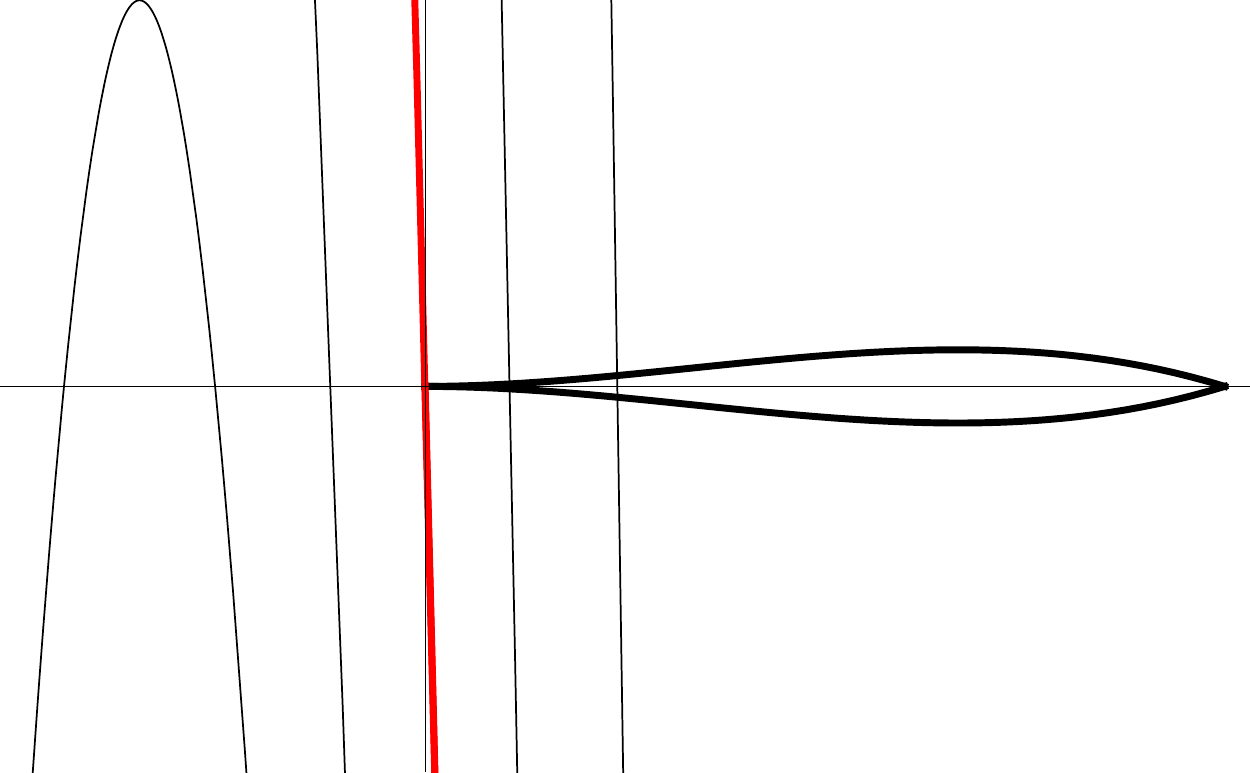}}
   \put(206,40){\footnotesize $u$}
   \put(60,113){\footnotesize $v$}
   \put(235,-10){\includegraphics[height=4.5cm,keepaspectratio]{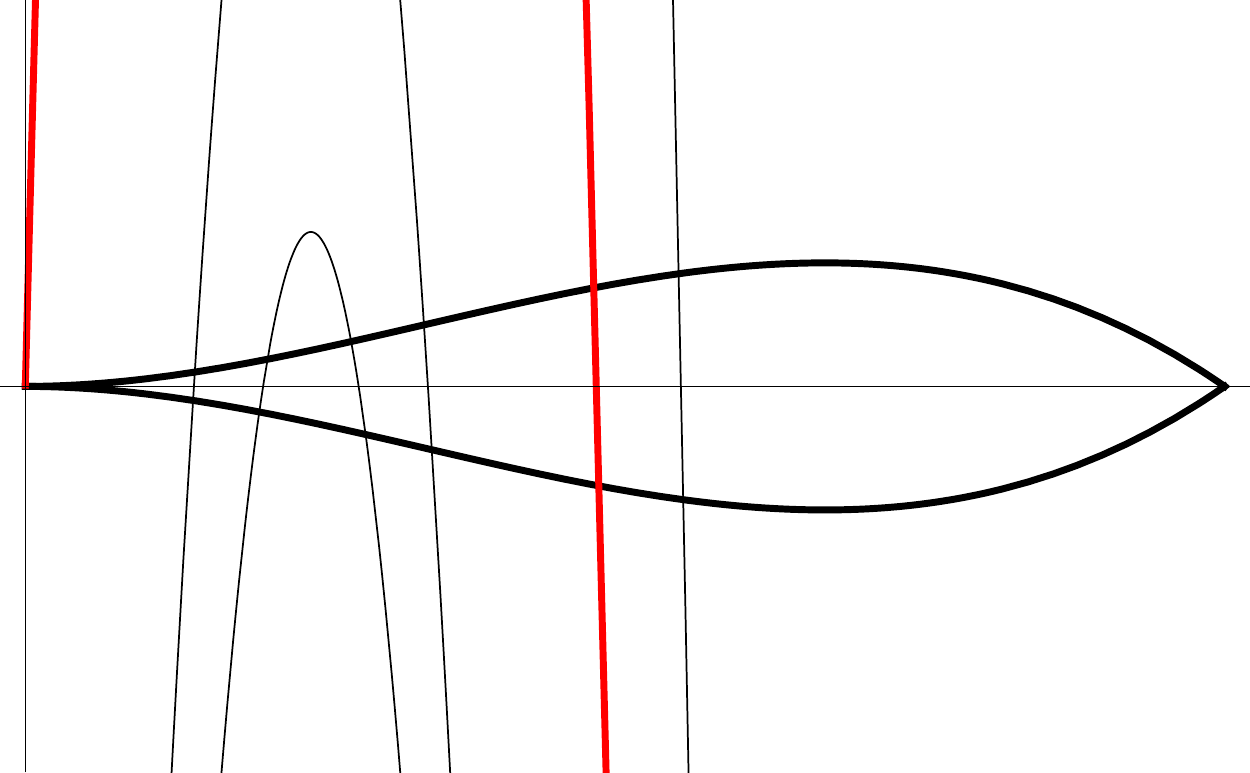}}
   \put(441,40){\footnotesize $u$}
   \put(244,113){\footnotesize $v$}
\end{picture}
\end{center}
\caption{\small{
Possible configurations between the (thick black) phase space
section $\cP^{\eta}\cap\{w=0\}$ and a second order approximation of
\eqref{generic curve} for $\delta=0.25$, $\alpha=-1$, $\lambda=0.35$,
$\mu=0.25$, $\varepsilon=0.2$ and $\eta=0.4$ (left), $\eta=0.6$ (right).
For values of $k$ corresponding to the  (thick) red curve we have a 
stable equilibrium at the origin (left) or a stable 
equilibrium at the origin and a periodic orbit around it 
(right). For values of $k$ slightly different (thin grey curves) 
we can have periodic orbits around the origin or no dynamics;  
in the right figure we furthermore have periodic orbits around
a regular equilibrium.
\label{fig3}}}
\end{figure}

At $\cQ_1 = (0, 0, 0)$ the reduced phase space section
$\cP^{\eta} \cap \{ w=0 \}$ has a cuspidal singularity. 
Suppose that~\eqref{generic curve} passes through the origin 
$(u, v) = (0, 0)$ with non-vanishing first derivative (see Fig.~\ref{fig3}).
Let us denote the corresponding value of $k$ by~$k_0$.
Recall that we assumed for definiteness that $\mu$ is positive and first
take $\lambda > 0$, so $\lambda \mu > 0$ and the derivative
of~\eqref{generic curve} at~$0$ is positive if $u_0 > 0$, compare with
Fig.~\ref{fig3}~(right).
Hence, values of~$k$ higher than~$k_0$ shift~\eqref{generic curve}
upward and correspond near~$\cQ_1$ to empty intersections of the
energy levels $\cK_{\delta, \varepsilon}^{\eta, h_0}(k)$ with the reduced
phase space~$\cP^{\eta}$ and thus to no dynamics.
Values of~$k$ lower than~$k_0$ shift~\eqref{generic curve}
downward and lead to periodic orbits around $\cQ_1$; in both cases
there may furthermore be periodic orbits where the second leaf of the
parabolic-cylinder-like level set~$\cK_{\delta, \varepsilon}^{\eta, h_0}(k)$
intersects~$\cP^{\eta}$, again compare with Fig.~\ref{fig3}~(right).
For $u_0 < 0$ values of $k$ higher than~$k_0$ yield periodic orbits (as
$v^{\prime}(0)$ is negative) and there are no additional intersections
for $k < k_0$, compare with Fig.~\ref{fig3}~(left).
The equilibrium~$\cQ_1$ is therefore stable for  $v^{\prime}(0) \neq 0$
and it can be unstable only if the curve~\eqref{generic curve} passes
through the origin $(u, v) = (0, 0)$ with vanishing first
derivative.
This happens for
\be
\label{eq: instability cusp}
v^{\prime}(0) \;\; = \;\;
- \frac{1}{\mu} \left[ \frac{2 \delta + \alpha \eta}{\varepsilon^2}
\, + \, \beta_2 \delta \eta \, + \, \gamma_3 \eta^2 \right]
\;\; = \;\; 0 \enspace .
\ee
Since we are following a perturbative approach, we look for a
solution of this equation in the form of a power series
$\eta = \eta_0 + \varepsilon^2 \eta_2$ in~$\varepsilon$.
For $\alpha \neq 0$ and $\delta \alpha \leq 0$ we obtain the critical
value
\be
\label{instability y}
\eta \;\; = \;\; \eta_1 \;\; := \;\; - \frac{2 \delta }{\alpha} \; + \;
\frac{2 \varepsilon^2 \delta ^2}{\alpha^3}(\beta_2 \alpha - 2 \gamma_3)
\enspace ,
\ee
with $\eta_0$ from~\eqref{etaatcusp} in the first order approximation.

\begin{remark}
This answers an open question from~\cite{MP2013a} where this critical
value for~$\eta$ was found with an ``empirical'' approach.
The \emph{two} families of periodic orbits, namely banana and
anti-banana orbits, bifurcate for the two-degree-of-freedom system
defined by the normal form and up to second order terms in the
perturbation this happens simultaneously, at the \emph{same} critical
value of~$\eta$.
Since $\cQ_1$ is a cusp point this has a geometric reason and
in particular subsists through all orders of the perturbation.
\end{remark}

\noindent
Note that 
\be
\label{second derivative}
v^{\prime \prime}(0) \;\; = \;\;
- \frac{2}{\mu} \left[ \frac{\lambda}{\varepsilon^2}
\, + \, \beta_1 \delta \, + \, \gamma_2 \eta \right]
\enspace ,
\ee
therefore we can assume that $v^{\prime \prime}(0)$ does not
vanish for small values of~$\varepsilon$, thus there is no degeneracy.
In case $\lambda < 0$ --- and hence $\lambda \mu < 0$ --- the
curve~\eqref{generic curve} has a minimum instead of a maximum
near~$u_0$, interchanging the effects of shifting~\eqref{generic curve}
upwards and downwards.
The above discussion applies {\it mutatis mutandis}, leading to the
same formula~\eqref{instability y} when $\alpha \neq 0$.

\begin{remark}
\label{remark:extra sol}
Equation~\eqref{eq: instability cusp} is of second order in~$\eta$,
therefore in general it admits two solutions for~$\eta$.
However, only one of these two solutions is convergent for
$\varepsilon \rightarrow 0$ and has a truncated series expansion
as in~\eqref{instability y}.
The second solution, once expressed as truncated power series 
of~$\varepsilon$, reads
$\eta = - \frac{\alpha}{\gamma_3 \varepsilon^2} +
 \left( \frac{2 \delta}{\alpha} - \frac{\beta_2}{\gamma_3} \right)$.
We aim at an approximation of the dynamics of the original
system~\eqref{Hamiltonian epsilon} in a neighbourhood of the origin
and at low energies.
Therefore here and in the computation of~\eqref{instability x} below we
disregard solutions that are divergent for $\varepsilon \rightarrow 0$
and limit the description of the dynamics only to low values of~$\eta$.
Note that $\eta$ can be related to the (generalized) energy in a
similar fashion as in~\eqref{energy on long axis}.
\end{remark}

\noindent
At $\cQ_2 = (\eta, 0, 0)$ the reduced phase space has a conical singularity.
The intersection of the reduced phase space~$\cP^{\eta}$ with the
$(u, v)$--plane is given by
\be
\label{reduced phase section}
\cC^{\eta}_{\pm} \;\; = \;\; \cP^{\eta} \; \cap \; \{ w = 0 \}
\;\; = \;\; \left\{\, (u, v) \in \R^2 \;:\;
v = \pm \frac{1}{2}(\eta-u) u^2 ,\; 0 \leq u \leq \eta \, \right\}
\ee
whence the slope of the two contour lines
constituting~\eqref{reduced phase section} at $(u, v) = (\eta, 0)$
is~$\mp \frac{1}{2} \eta^2$.
By the same argument we used above, the corresponding equilibrium
can be unstable only if the slope of the curve~\eqref{generic curve}
at $(u, v) = (\eta, 0)$ takes values in the interval
$\opin{{-} \frac{1}{2} \eta^2}{\frac{1}{2} \eta^2}$.
Thus, to find the critical values for~$\eta$ which correspond
to stability/instability transitions of the equilibrium, we need to
solve
\be
\label{eq_etapm}
v^{\prime}(\eta) \;\; = \;\; \mp \frac{\eta^2}{2} \enspace .
\ee
Proceeding as before, we look for solutions of the form
$\eta = \eta_0 + \varepsilon^2 \eta_2$ with $\eta_0$ from~\eqref{etaatcone}.
We arrive at the two solutions $\eta = \eta_{\pm}$ given by
\be
\label{instability x}
 \eta_{\pm} \;\; := \;\;
\frac{- 2 \delta }{\alpha + 2 \lambda } \; + \;
\frac{2 \varepsilon^2 \delta^2}{(\alpha + 2 \lambda)^3}
\, \left[ 2 \beta_1 (\alpha + 2 \lambda) \, + \,
\beta_2 (\alpha + 2 \lambda ) \, - \, (\gamma \pm \mu) \right]
\enspace ,
\ee
where $\gamma = 6 \gamma_1 + 4 \gamma_2 + 2 \gamma_3$.
Such solutions are acceptable if $\alpha + 2 \lambda \neq 0$ and
$\delta (\alpha + 2 \lambda) \leq 0$.
Since 
\bd 
v''(\eta) \;\; = \;\; - \frac{2}{\mu} \left[ \frac{\lambda}{\varepsilon^2}
\, + \, \beta_1 \delta \, + \, (3 \gamma_1 + \gamma_2) \eta \right]
\ed
and $\lambda \neq 0$ we can (as in~\eqref{second derivative}) 
conclude that there is no degeneracy. 
Note that the difference between the threshold values
in~\eqref{instability x} is
\be
\label{eta difference}
\eta_- \; - \; \eta_+ \;\; = \;\;
\frac{4 \varepsilon^2 \delta ^2 \mu }{(\alpha + 2 \lambda )^3}
\enspace.
\ee
Therefore, the equilibrium is
unstable for $\eta_- < \eta < \eta_+$ if $\alpha + 2 \lambda < 0$ and
for $\eta_+ < \eta < \eta_-$ if $\alpha + 2 \lambda > 0$.
For $\mu < 0$ it is the other way around.

\subsection{Regular equilibria}
\label{sec:regularequilibria}

Regular equilibria correspond to points where the level
sets~$\cK_{\delta, \varepsilon}^{\eta, h_0}(k)$
touch (i.e.\ are tangent to) the reduced phase space~$\cP^{\eta}$.
The normal form~\eqref{reduced Hamiltonian w} is independent
of the variable~$w$, whence the level
sets~$\cK_{\delta, \varepsilon}^{\eta, h_0}(k)$ are cylinders
(consisting of lines parallel to the $w$--axis) on the basis of
the curve~\eqref{generic curve}.
However, a tangent plane to the surface of revolution~$\cP^{\eta}$
can contain the $w$--axis only at points $(u, v, w)$ with $w = 0$.
Thus, $\cK_{\delta, \varepsilon}^{\eta, h_0}(k)$ and $\cP^{\eta}$
can touch each other only at points in the $(u, v)$--plane; this
is what we achieved when rotating to $\nu = 0$.
The intersection of~$\cP^{\eta}$ with the $(u, v)$--plane is
given by~\eqref{reduced phase section} whence regular equilibria
correspond to the points $u \in \opin{0}{\eta}$ in which
\eqref{generic curve} and~\eqref{reduced phase section} intersect
with coinciding slopes.
As we can always adjust the second order part~$k$ of the energy
to make \eqref{generic curve} and~\eqref{reduced phase section}
intersect where desired, this gives the equation
\bd
2 \delta + \alpha \eta \, + \, 2 \lambda u \; - \;
\varepsilon^2 \left[ 3 (2 \gamma_1 \pm \mu) \frac{u^2}{2} \, + \,
[2 \delta \beta_1 + \eta (2 \gamma_2 \mp \mu)] u \, + \,
\delta \beta_2 \eta + \gamma_3 \eta^2 \right] \;\; = \;\; 0
\ed
for the slopes to coincide.
Looking for a solution of the form $u = u_0 + \varepsilon^2 u_2$,
we find the two solutions
\bd
u \;\; = \;\; u_{\pm} \;\; := \;\;
u_0 \; + \; \varepsilon^2 u_2^{\pm}
\ed
subject to $0 \leq u_{\pm} \leq \eta$, where
\bd
u_2^{\pm} \;\; = \;\; \frac{4 \lambda (2 \delta + \alpha \eta)
[2 \delta \beta_1 + \eta (2 \gamma_2 \mp \mu)] \, - \,
8 \eta \lambda ^2 (\delta \beta_2 + \gamma_3 \eta) \, - \,
3 (2 \gamma_1 \pm \mu)(2 \delta + \alpha \eta)^2}
{16 \lambda ^3}
\ed
and $u_0$ as in~\eqref{zero order tangency}.
Solving $u_{\pm} = \eta$ for $\eta = \eta_0 + \varepsilon^2 \eta_2$
we recover~\eqref{instability x}, while solving $u_{\pm} = 0$ we
recover~\eqref{instability y}.

Therefore, as expected, the bifurcation of regular equilibria is
related to the transition to instability of singular equilibria.
Since at $\cQ_1 = (0, 0, 0)$ there is a cusp singularity, the
corresponding equilibrium is unstable only at $\eta = \eta_1$.
However at this critical value, \emph{two} tangency points
appear/disappear simultaneously and therefore \emph{two} regular
equilibria bifurcate off from the origin.
On~$\cP^{\eta}$ they correspond to two points
$\cU_{\pm} = (u_{\pm}, v_{\pm}, 0)$ with
\bd
v_{+} \; = \; - \frac{1}{2} (\eta - u_{+}) u_{+}^2
\quad \mbox{and} \quad
v_{-} \; = \; + \frac{1}{2} (\eta - u_{-}) u_{-}^2
\enspace ,
\ed
one lying on the lower and one on the upper contour of the reduced
phase space.
The singular equilibrium~$\cQ_2 = (\eta, 0, 0)$, on the other
hand, can change its stability twice, since it corresponds to a
conical singularity.
Each stability/instability transition is associated with the
appearance or disappearance of only one regular equilibrium.
The bifurcating equilibria correspond to the point~$\cU_-$
on the upper contour of the reduced phase space for
$\eta = \eta_-$ and to the point~$\cU_+$ on the lower
contour for $\eta = \eta_+$.
We discuss the implications for the original system in
section~\ref{sec:bifurcations}.

\begin{figure}[htb]
\begin{center}
\begin{picture}(400,230)
   \put(0,114){\includegraphics[height=4cm,keepaspectratio]{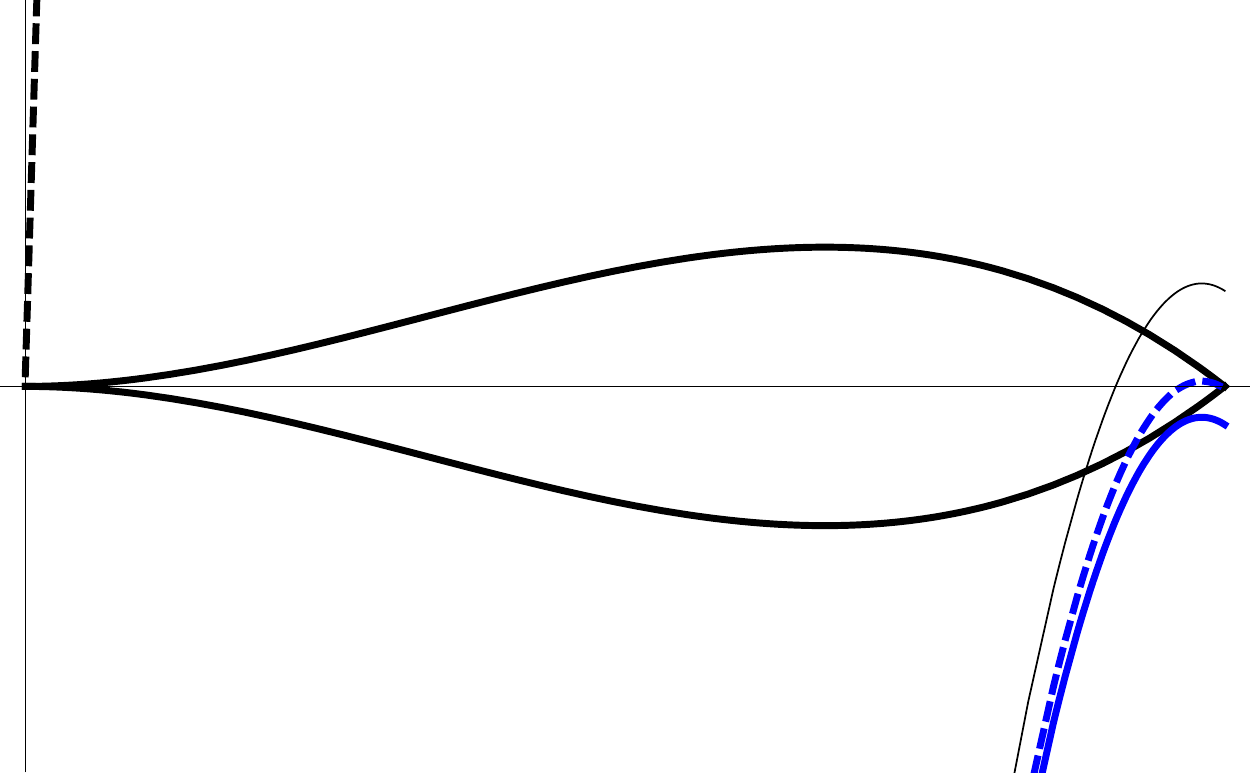}}
   \put(178,159){\footnotesize $u$}
   \put(8,224){\footnotesize $v$}
   \put(213,114){\includegraphics[height=4cm,keepaspectratio]{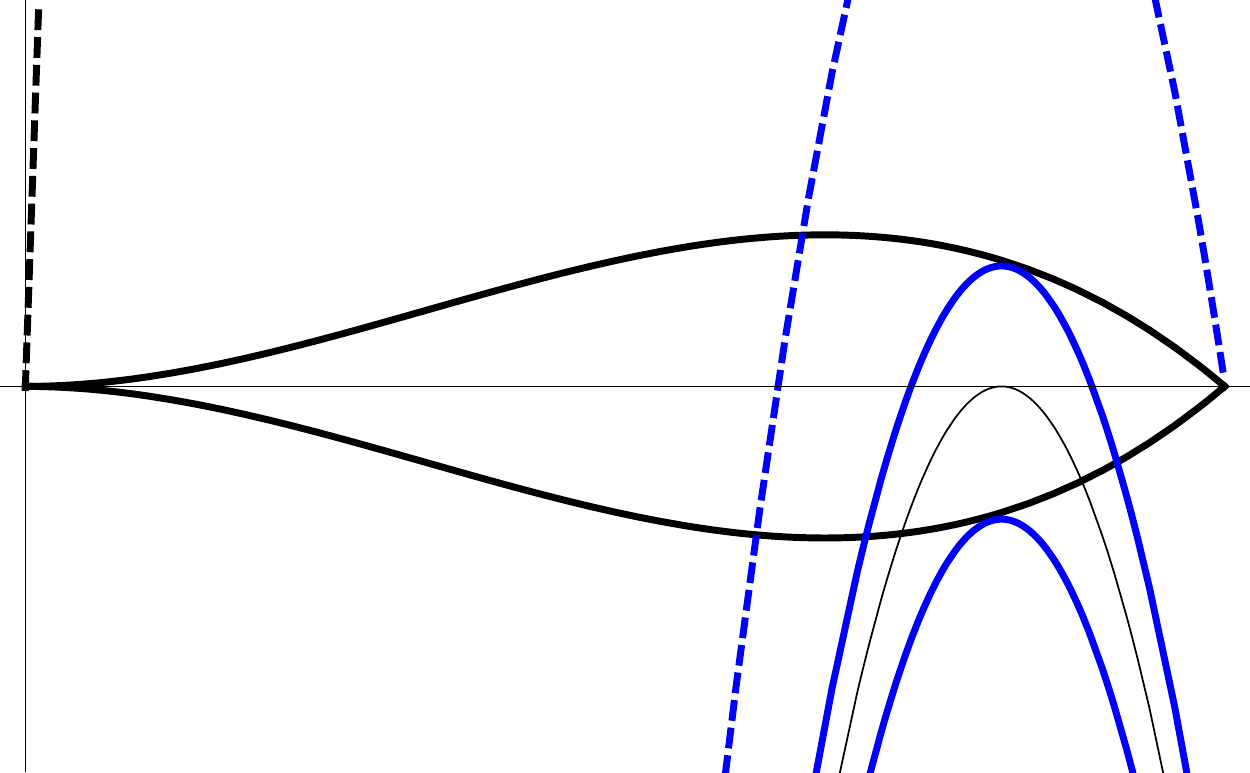}}
   \put(391,159){\footnotesize $u$}
   \put(221,224){\footnotesize $v$}
   \put(0,-10){\includegraphics[height=4cm,keepaspectratio]{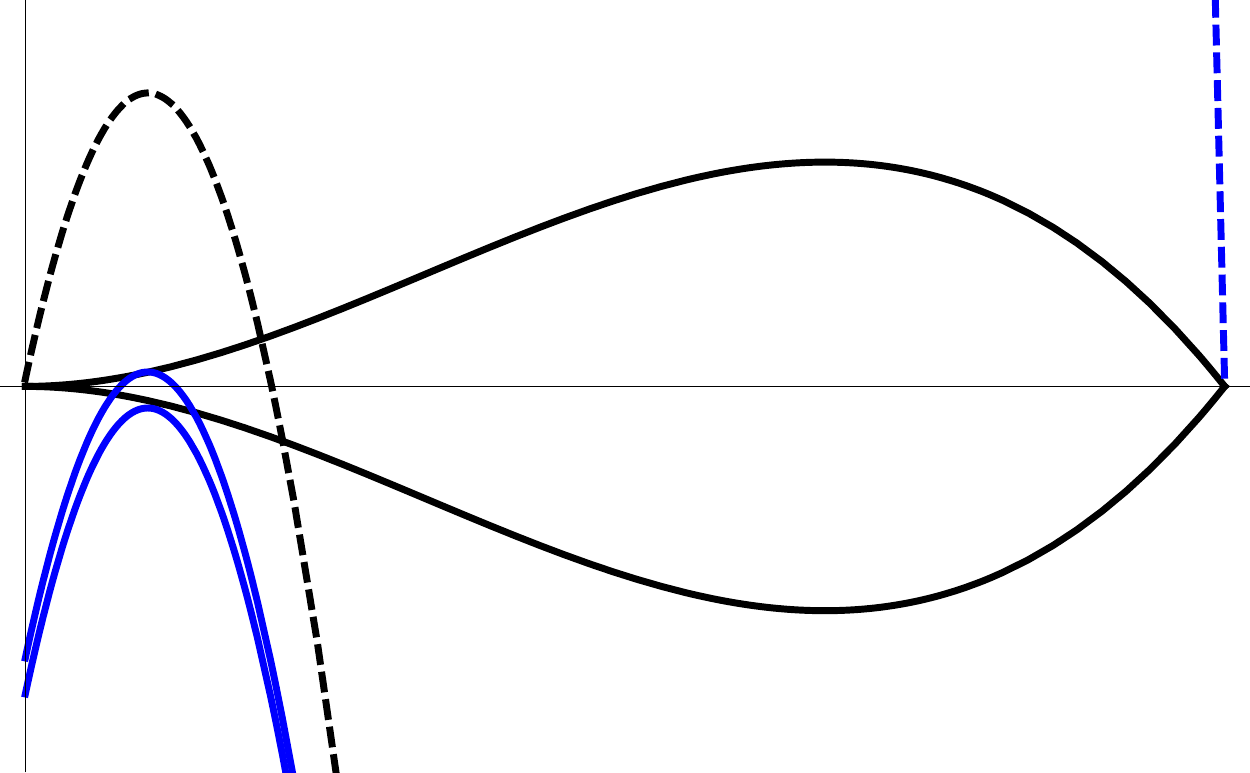}}
   \put(178,35){\footnotesize $u$}
   \put(8,100){\footnotesize $v$}
   \put(213,-10){\includegraphics[height=4cm,keepaspectratio]{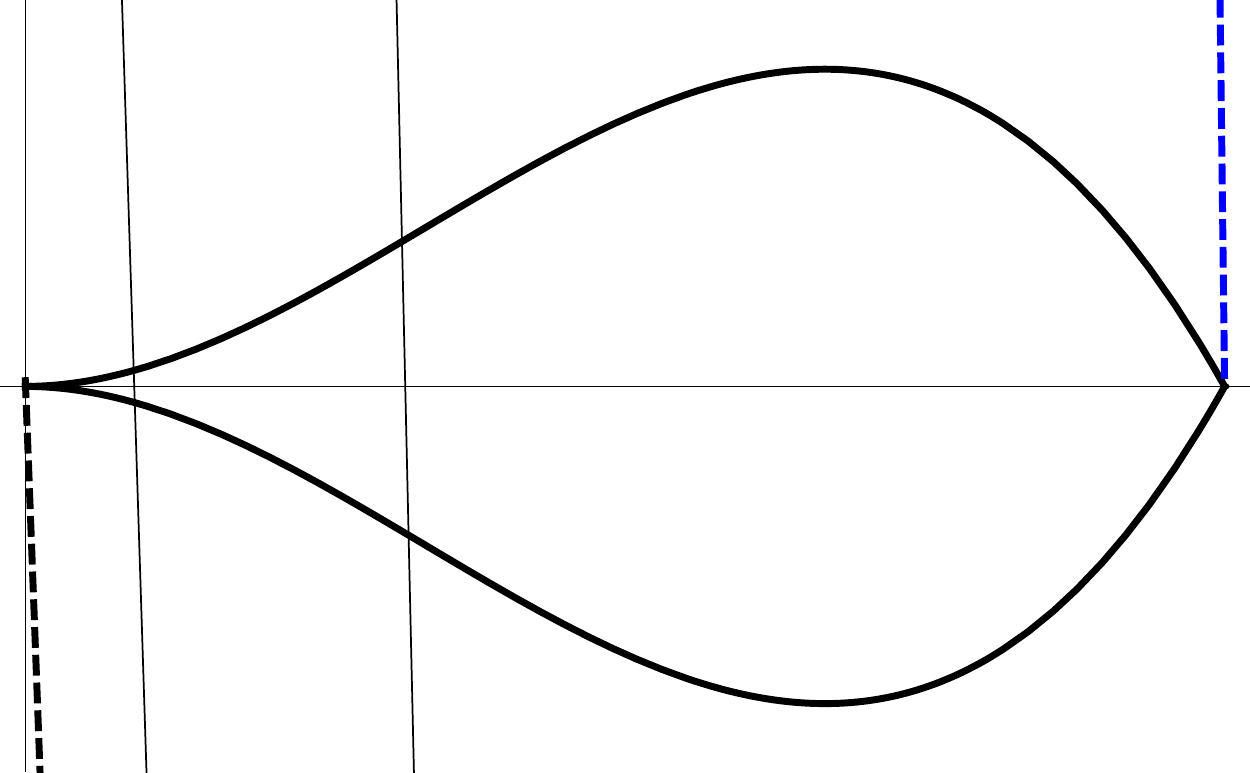}}
   \put(391,35){\footnotesize $u$}
   \put(221,100){\footnotesize $v$}
\end{picture}
\end{center}
\caption{\small{
Possible tangencies between the parabola~\eqref{parabola} and the
phase space section $\cP^{\eta} \cap \{ w = 0 \}$
in~\eqref{reduced phase section} for increasing
values of~$\eta$ and fixed values $\delta = -0.25$, $\mu = 0.25$,
$\lambda = 0.1$, $\alpha = 1$ and $\varepsilon = 0.3$
of the detuning and the other parameters.
Two regular equilibria appear successively from the conical singularity
and subsequently disappear simultaneously on the singular equilibrium at
the origin.
The equilibrium on the upper contour of the phase space is unstable while
the equilibrium on the lower contour is stable.
Upper left: $\eta = 0.418$.
First a stable equilibrium appears from the singular equilibrium
$\cQ_2 = (\eta, 0, 0)$, which as a consequence becomes unstable.
Upper right: $\eta = 0.43$.
Then an unstable equilibrium appears from the singular
equilibrium~$\cQ_2$, which becomes stable after the bifurcation.
Lower left: $\eta = 0.49$.
Both regular equilibria are going to disappear when reaching the stable
singular equilibrium $\cQ_1 = (0, 0, 0)$ at the cusp singularity.
Lower right: $\eta = 0.55$.
Both regular equilibria disappeared and the only equilibria are the
singular ones, both stable.
}}\label{Fig: X2ab} 
\end{figure}
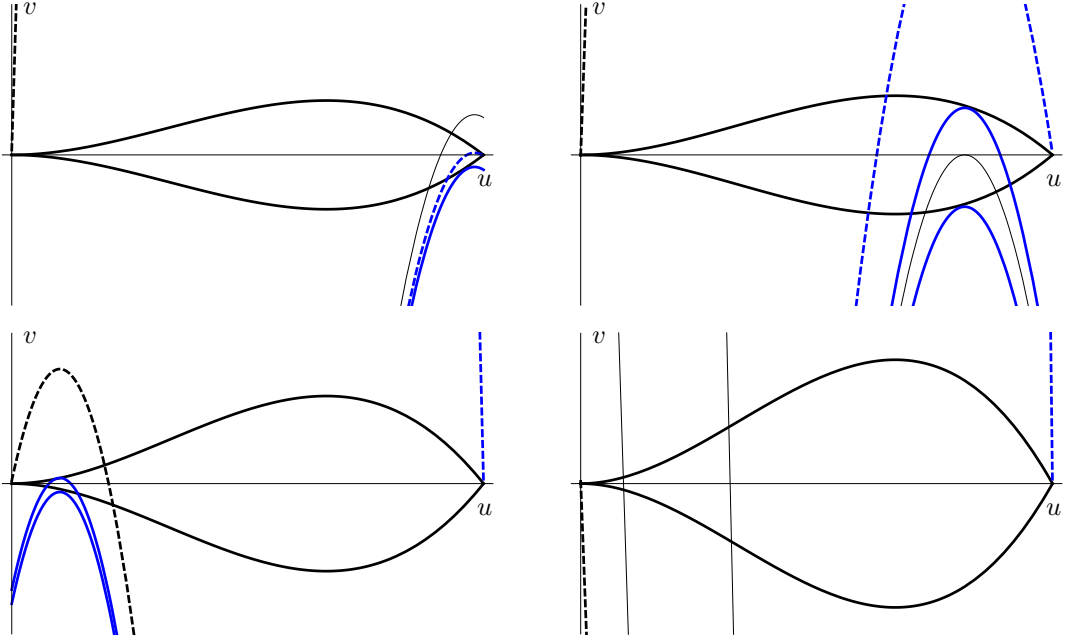

Let us conclude this section with the analysis of the
stability of the regular equilibria.
Once we know that the two curves \eqref{generic curve}
and~\eqref{reduced phase section} touch, to study the stability of
the corresponding equilibrium we need to know ``how'' they touch.
Indeed, for small value of~$\varepsilon$, the curvature
of~\eqref{generic curve} is determined by its second order
approximation given by the parabola~$P$ defined by
\be
\label{parabola}
v \;\; = \;\; P(u) \;\; := \;\; \frac{1}{\mu} \left[
k \, - \, \frac{\lambda}{\varepsilon^2}(u - u_0)^2
\right] \enspace .
\ee
Moreover, the smaller the value of~$\varepsilon$, the greater in
absolute value the curvature of such a parabola.
Along the limit $\varepsilon \to 0$ the curvature
of~\eqref{generic curve} can always be set greater in absolute
value than the curvature of the contour of the reduced phase space
at a tangency point. 

To fix the ideas, let us consider the case when the parabola
touches the phase space section at its lower arc~$\cC^{\eta}_-$
``from outside'', i.e.\ there is no intersection point other
than the tangency point.
Let $k_+$ be the corresponding level for~$k$.
We can always assume the curvature of~$P$ to be high enough (in absolute
value) so that this can happen only if the parabola is upside-down, i.e.\
concave; as $\mu > 0$ this is equivalent to $\lambda > 0$.
Higher values of~$k_+$ then shift the parabola upward and correspond
to closed orbits around the equilibrium --- which therefore is stable
(see Fig.~\ref{Fig: X2ab} upper right) --- until the maximum of~$P$
reaches the upper contour.

If the parabola~\eqref{parabola} is convex, then it touches the lower
contour of the phase space ``from inside'', i.e.\ there are two
further intersections on the upper contour of the phase space.
In this case the equilibrium is unstable.
This happens for $\lambda < 0$ and qualitatively amounts to flipping
Fig.~\ref{Fig: X2ab} upside down.
The stable and unstable manifolds of the equilibrium are determined
by the intersection curves between the reduced phase
space~$\cP^{\eta}$ and the energy level set
$\cK_{\delta, \varepsilon}^{\eta, h_0}(k_+)$, i.e.\ the surface
corresponding to~\eqref{generic curve} for
$h = h_0 + \varepsilon^2 k_+ $. 

Similarly, for the stability analysis of the equilibrium on the
upper contour of the phase space, we find that it is stable for
$\lambda < 0$ and unstable for $\lambda > 0$; for $\mu < 0$ it
is the other way around.

\begin{remark}
The simple geometry of the parabola allows to immediately
conclude stability or instability of the equilibria and how these
come into existence through centre-saddle and Hamiltonian flip
bifurcations.
The corresponding formulas may as well be searched for as double
roots of the difference of the polynomials describing phase space
and energy level set~\cite{EHM18}.
For more involved expressions than the present cubic, which is well
approximated by a parabola, an algebraic point of view can support
the present geometric approach, relying on the resultant of two
polynomials and related tools.
\end{remark}

\subsection{The completed bifurcation diagram}
\label{sec:completed bifurcation diagram}

The dissolution of the great circle of equilibria into a stable
and an unstable equilibrium, with a family of periodic orbits inside
the separatrix of the latter surrounding the former, allows to
complete the bifurcation diagram obtained in
section~\ref{sec:bifurcation diagram}.
Indeed, the region~$\cD_{\lambda}^{\alpha}$ --- the green sector in
Fig.~\ref{fig2} --- no longer stands for structurally unstable
dynamics.
The blue line in Fig.~\ref{fig2} splits into the two lines
$\eta_{-} = \eta_{-}(\delta)$ and $\eta_{+} = \eta_{+}(\delta)$.
 Between these lines the dynamics is as depicted in
Fig.~\ref{Fig: X2ab} (upper right/lower left).
The other boundary line of~$\cD_{\lambda}^{\alpha}$ --- the red line
in Fig.~\ref{fig2} --- does not split but gets refined from
\eqref{etaatcusp} to~\eqref{instability y} and now stands for the
simultaneous bifurcation at $\cQ_1 = (0, 0, 0)$ where the two regular
equilibria disappear into the singular equilibrium~$\cQ_1$.

\begin{figure}[p]
\begin{center}
\begin{picture}(450,450)
  \put(0,366){\includegraphics[height=3cm,keepaspectratio]{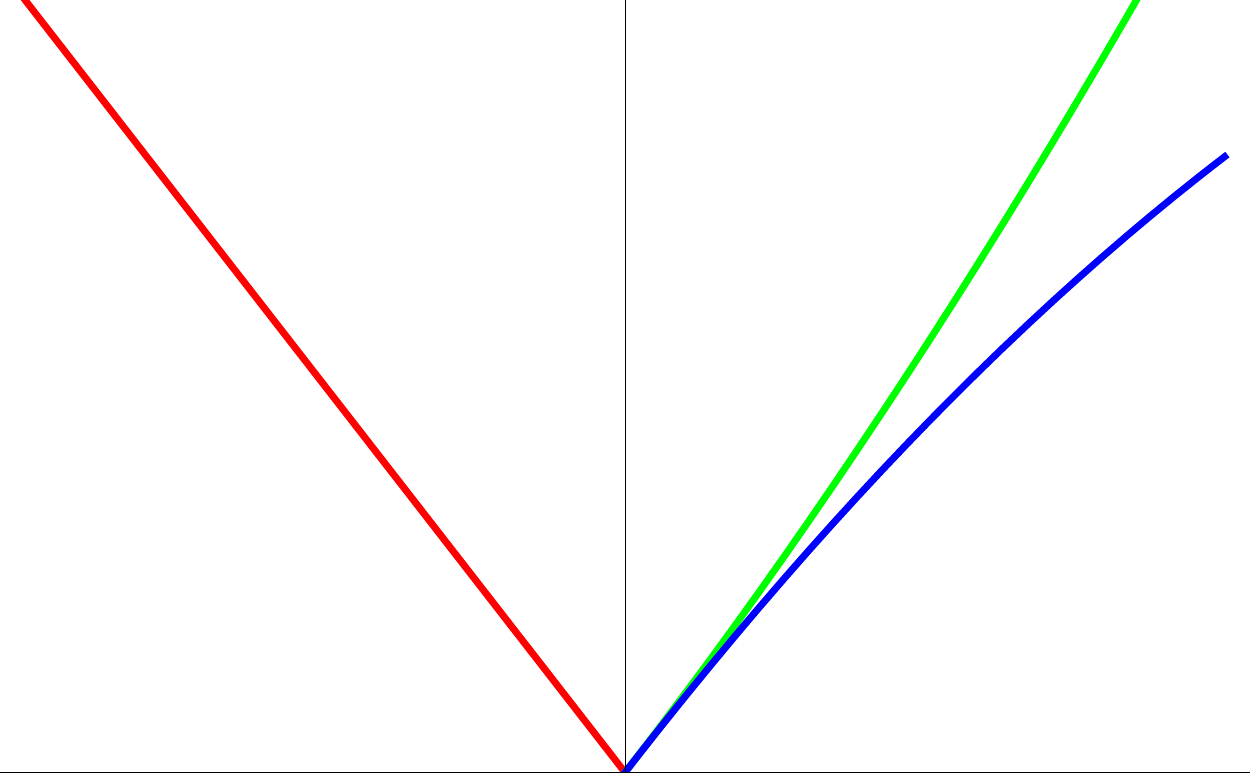}}
  \put(137,368){\footnotesize $\delta$}
  \put(71,446){\footnotesize $\eta$}
  \put(157,366){\includegraphics[height=3cm,keepaspectratio]{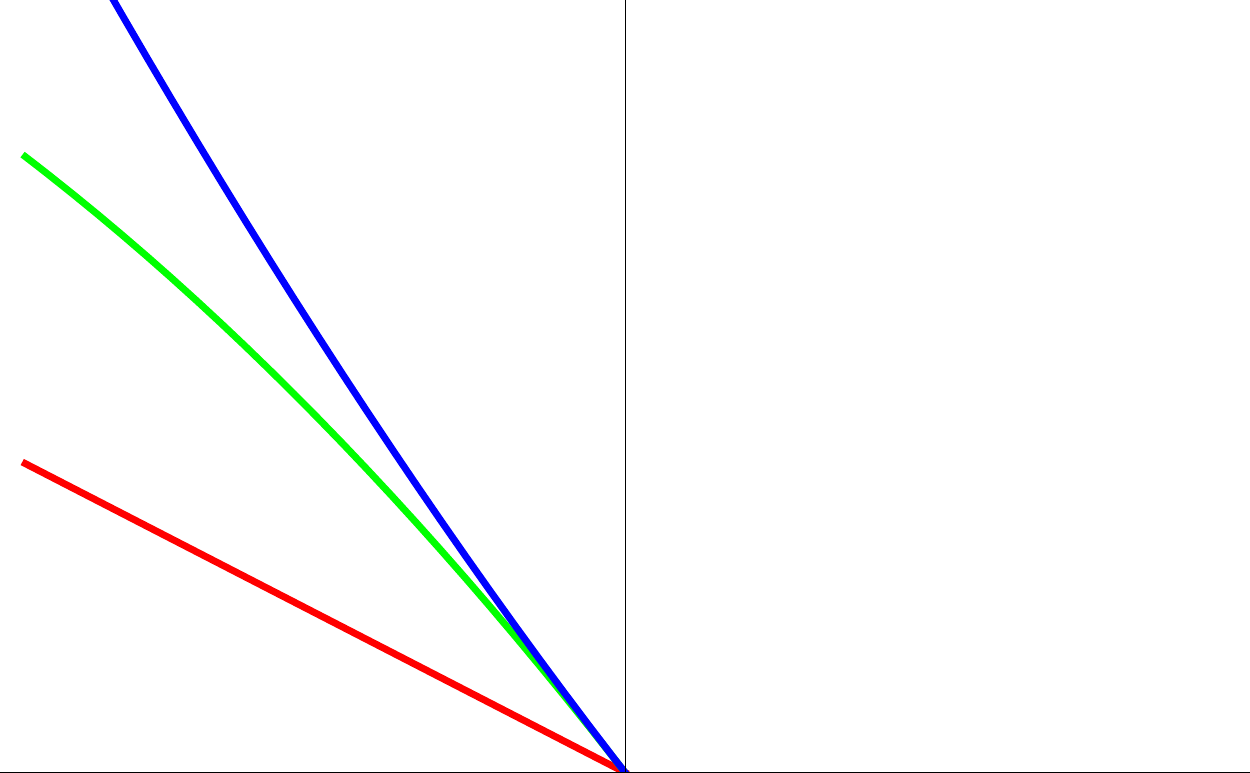}}
  \put(294,368){\footnotesize $\delta$}
  \put(228,446){\footnotesize $\eta$}
  \put(316,366){\includegraphics[height=3cm,keepaspectratio]{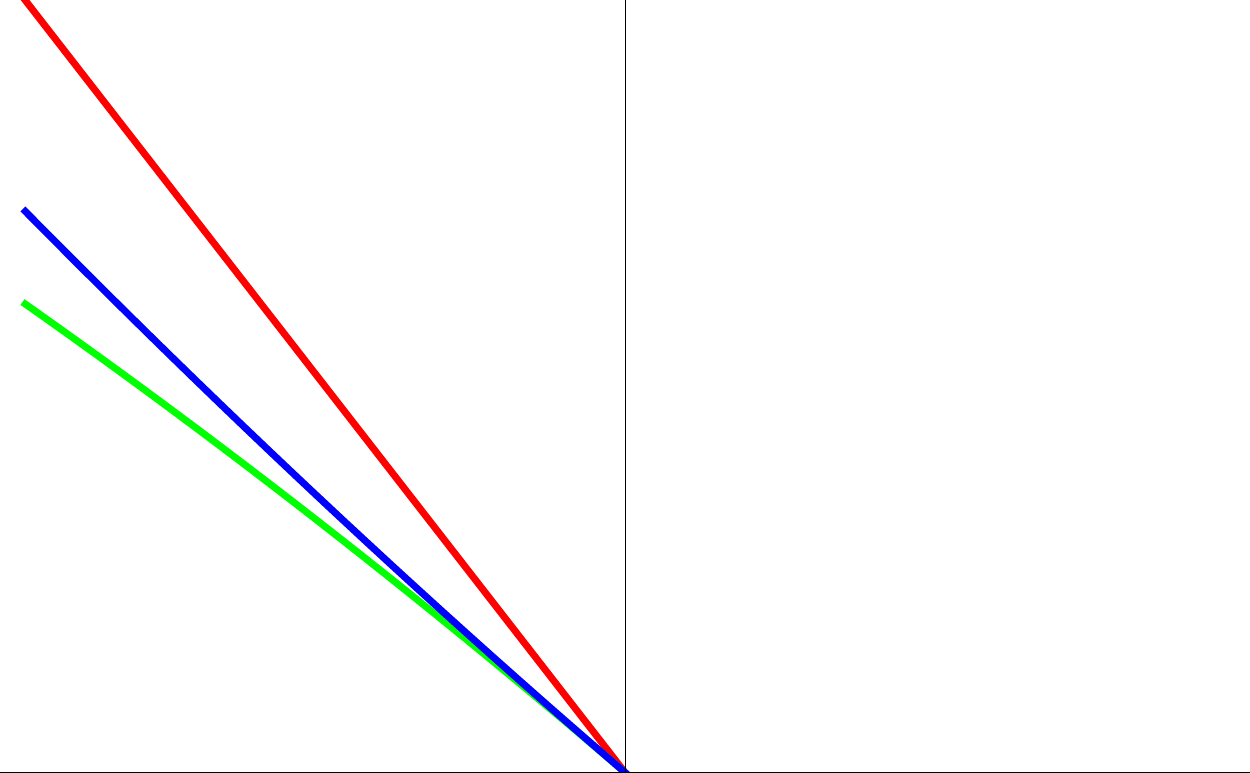}}
  \put(453,368){\footnotesize $\delta$}
  \put(387,446){\footnotesize $\eta$}
  \put(112,112){\includegraphics[height=8cm,keepaspectratio]{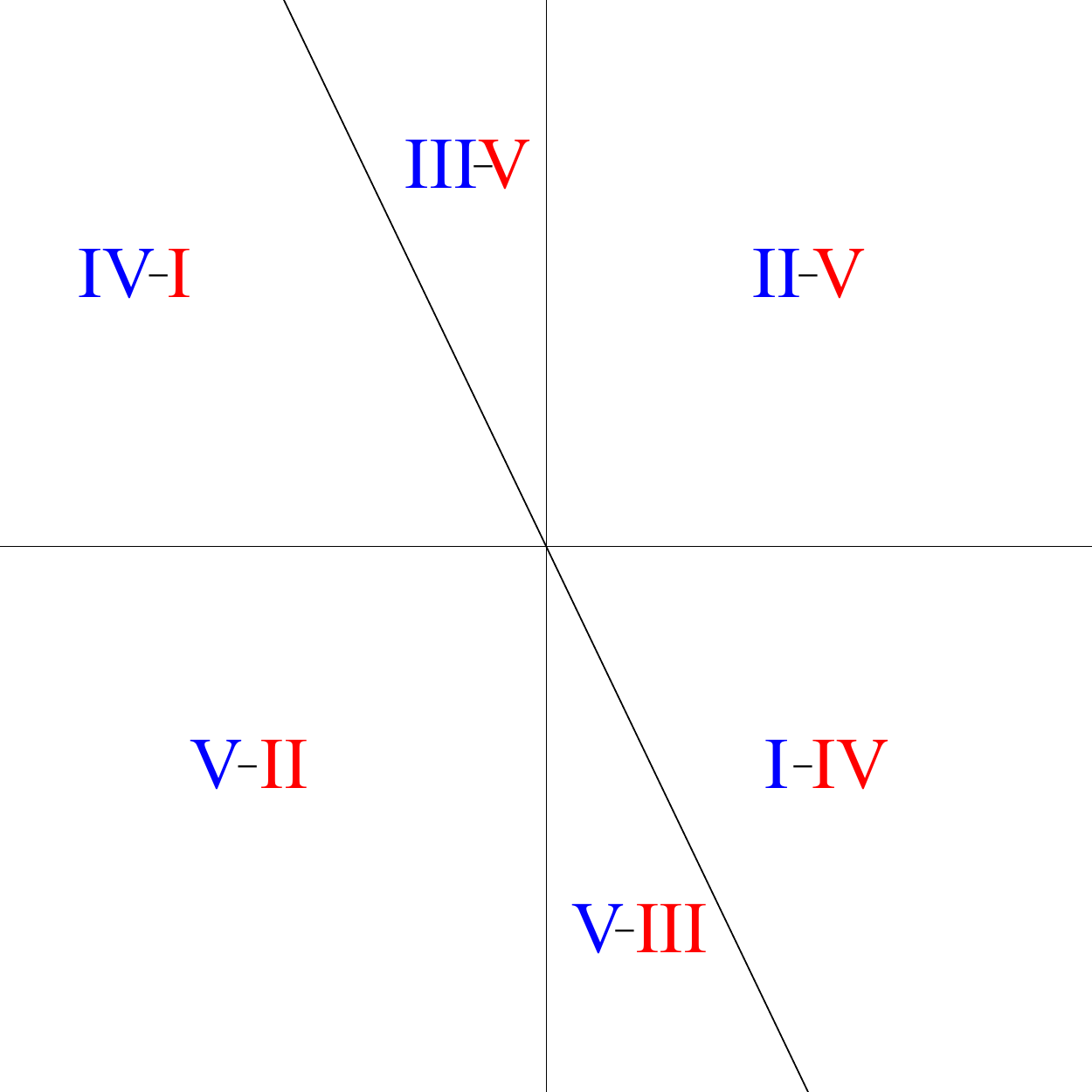}}
  \put(346,222){\footnotesize $\lambda$}
  \put(232,336){\footnotesize $\alpha$}
  \put(0,0){\includegraphics[height=3cm,keepaspectratio]{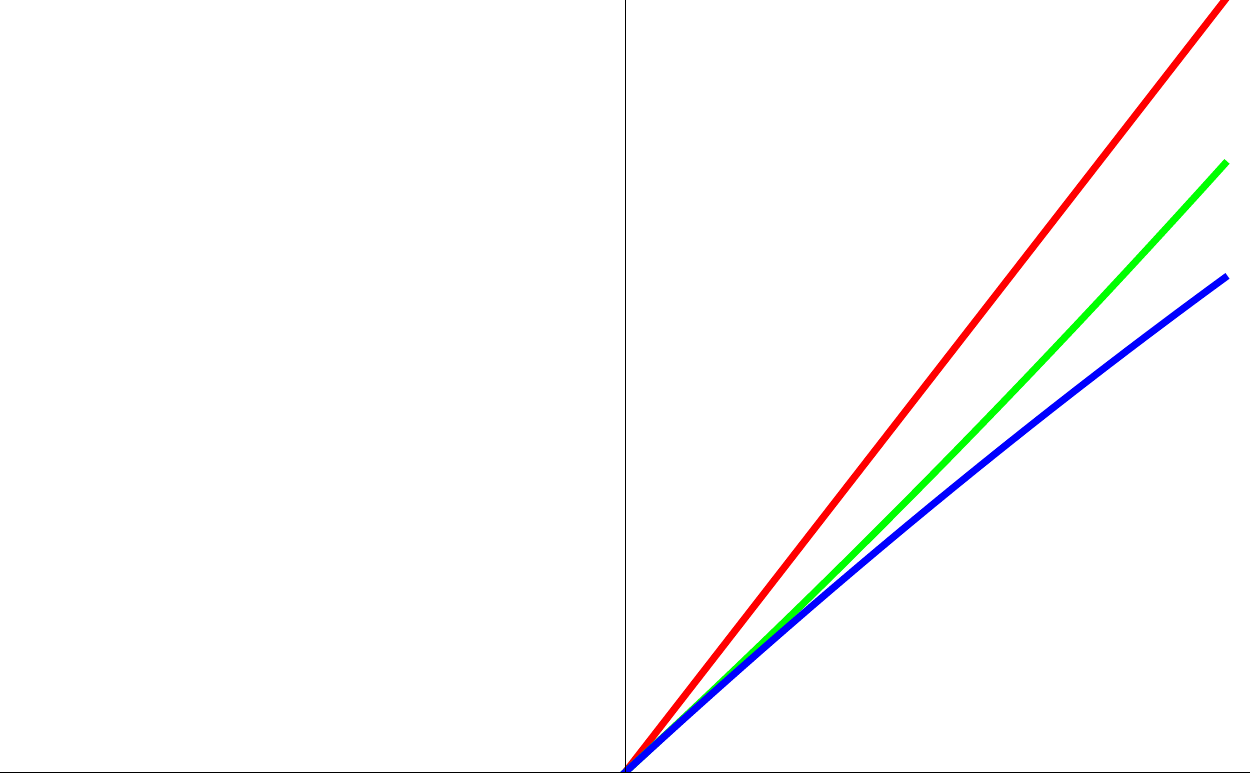}}
  \put(137,2){\footnotesize $\delta$}
  \put(61,80){\footnotesize $\eta$}
 \put(157,0){\includegraphics[height=3cm,keepaspectratio]{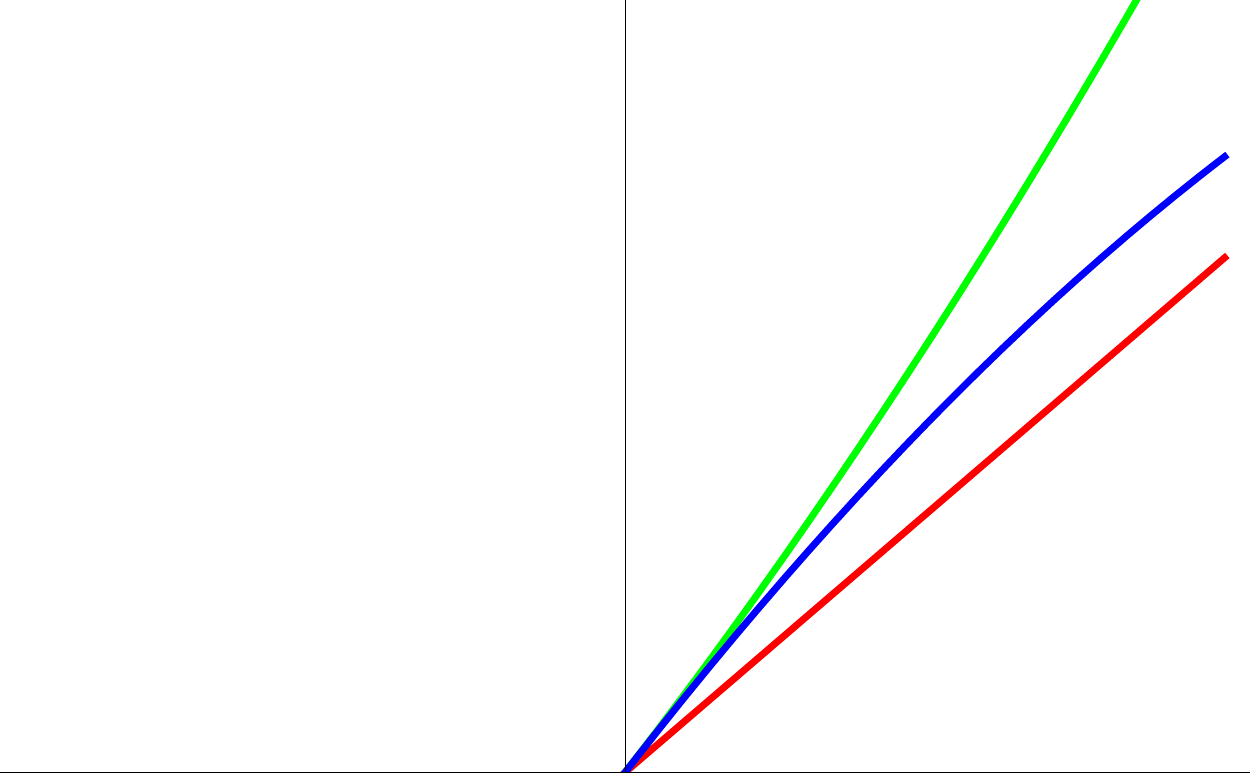}}
  \put(294,2){\footnotesize $\delta$}
  \put(218,80){\footnotesize $\eta$}
  \put(316,0){\includegraphics[height=3cm,keepaspectratio]{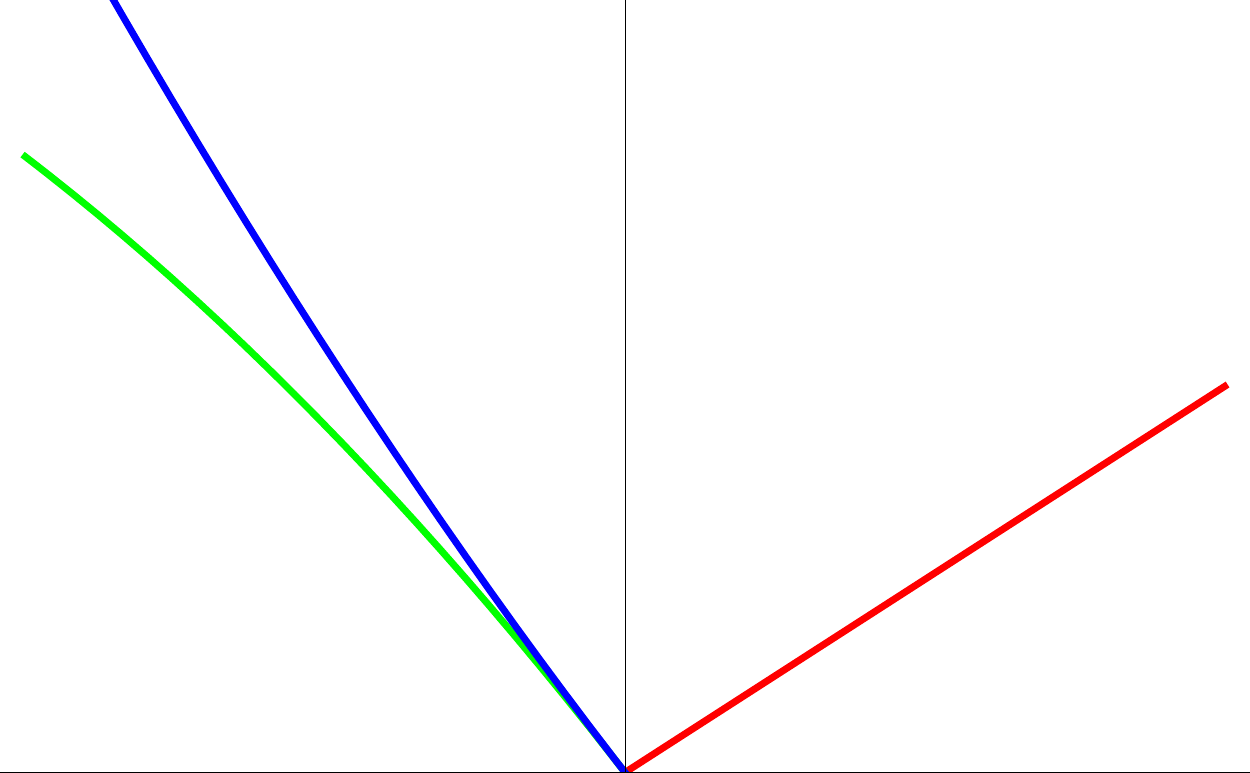}}
  \put(453,2){\footnotesize $\delta$}
  \put(375,80){\footnotesize $\eta$}
\end{picture}
\end{center}
\caption{\small{
Middle: diagram showing on the $(\lambda, \alpha)$--plane the 
regions corresponding to Cases~I--V, in blue for $\delta < 0$ and in 
red for $\delta > 0$.
The black line gives the boundary $\alpha + 2 \lambda = 0$.
Clockwise around: 
bifurcation diagrams on the $(\delta,\eta)$--plane corresponding to the
various cases (Case~\blue{IV}-\red{I} in the right above,
Case~\blue{III}-\red{V} in the middle above, etc.).
The bifurcation thresholds $\eta_1$, $\eta_-$ and $\eta_+$ 
are in red, blue and green, respectively.
}}\label{fig: daDiagram}
\end{figure}
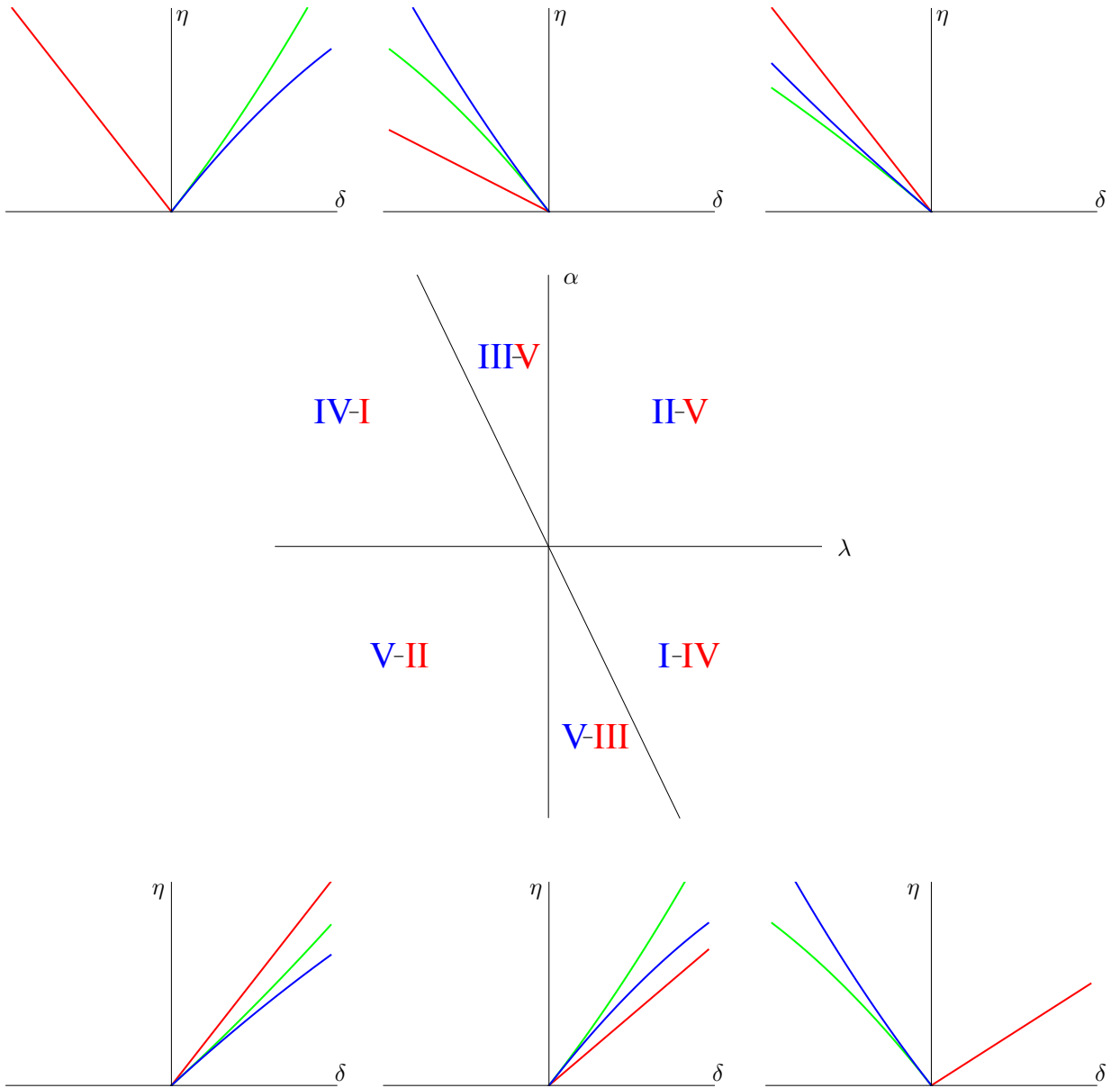

The resulting possibilities are assembled in Fig.~\ref{fig: daDiagram}.
The central $(\lambda, \alpha)$--plane allows to distinguish the six
cases \blue{I}-\red{IV} to \blue{V}-\red{III} --- when passing through
one of the lines $\{ \alpha = 0 \}$, $\{ \lambda = 0 \}$ and
$\{ \alpha + 2 \lambda = 0 \}$ the bifurcation diagram in the
$(\delta, \eta)$--plane changes.
The bifurcation diagrams for $\alpha + 2 \lambda > 0$ and
$\alpha + 2 \lambda < 0$ are related through a reflection with respect
to the $\eta$--axis, together with exchanging the blue and green
thresholds.
Varying $\delta$ through~$0$ yields a passage through resonance, for
reasonably small values of $\eta \geq 0$ near $\delta = 0$.
Note that since we reduced the dynamics on~$\cP^{\eta}$
through~\eqref{Z coordinates}, every regular equilibrium
on~$\cP^{\eta}$ corresponds to two regular equilibria
on~$\cV^{\eta}$.
Namely, the equilibrium~$\cU_-$ gives two
equilibria $\cA_{\pm}$ on~$\cV^{\eta}$.
Such equilibria lie on the intersection between $\cV^{\eta}$ and
the plane $\sigma_2 = 0$ and are symmetric with respect to the
plane $\sigma_1 = 0$.
These reconstruct in two degrees of freedom to the anti-banana orbits.
Similarly, the equilibrium~$\cU_+$ corresponds to two
equilibria $\cB_{\pm}$ on $V^{\eta} \cap \{ \sigma_1 = 0 \}$,
symmetric with respect to the plane $\sigma_2 = 0$.
From these the banana orbits are reconstructed in two degrees
of freedom.

\subsection{The bifurcation sequences}
\label{sec:bifurcation sequences}

In the previous section we have treated the detuning~$\delta$ as a
parameter.
However, the value $\eta \geq 0$ of the integral~$H_0$ is a
distinguished parameter with respect to~$\delta$ and one can in fact
consider the three coefficients $\delta$, $\alpha \neq 0$ and
$\lambda \neq 0$ in~\eqref{reduced Hamiltonian w} as fixed with
$\alpha + 2 \lambda \neq 0$, take $\mu > 0$ (think of
$\mu = 1$, although we refrain from explicitly performing this
re-parametrisation) and ignore the values of
$\beta_1, \beta_2, \gamma_1, \gamma_2, \gamma_3$
in~\eqref{newK4} which --- for sufficiently small values of
$\delta$ and~$\eta$, cf.\ remark~\ref{remark:extra sol}
--- do not change the dynamics.
Varying $\eta$ then yields the bifurcation sequence.
While the signs of $\alpha$ and~$\alpha + 2 \lambda$ determine
the sign of the first order approximation of \eqref{instability y}
and~\eqref{instability x}, the sign of $\alpha + 2 \lambda$ also
decides the bifurcation order of $\cU_+$ and~$\cU_-$ in force
of~\eqref{eta difference}.
To fix the ideas, let us start by assuming $\delta < 0$.
The five possible cases \blue{I--V}, which are described in the following,
are in accordance with the labeling in Fig.~\ref{fig: daDiagram}.

\begin{description}
\item{Case \blue{I}. $\alpha < 0 < \alpha + 2 \lambda$.}
In this case the critical value~$\eta_1$ is not acceptable: the
red line in Fig.~\ref{fig: daDiagram} (lower-right) does not pass
through the left quadrant.
Bifurcations can occur only when $\eta$ passes through the critical
values~$\eta_{\pm}$, with $\eta_+ < \eta_-$.
Since $\lambda \mu > 0$ the parabola~\eqref{parabola} is concave.
From section~\ref{sec:regularequilibria} we easily see
that at $\eta = \eta_+$ a stable equilibrium~$\cU_+$ appears from
the singular equilibrium~$\cQ_2$ that becomes unstable.
At $\eta = \eta_-$ the singular equilibrium~$\cQ_2$ turns stable
again and an unstable equilibrium~$\cU_-$ appears.
The equilibrium at~$\cQ_1$ always stays stable.
Increasing $\eta$ beyond~$\eta_-$ does increase the size
of~$\cP^{\eta}$, but the configuration of equilibria remains
qualitatively that of Fig.~\ref{Fig: X2ab} (upper-right and lower-left).

\item{Case \blue{II}. $0 < \alpha < \alpha + 2 \lambda$.}
All critical values are positive now, with $\eta_+ < \eta_- < \eta_1$.
The parabola~$P$ is still concave.
As in the previous case, we see first the appareance of one stable
equilibrium~$\cU_+$ at $\eta = \eta_+$, while $\cQ_2$ becomes unstable.
Then, an unstable equilibrium~$\cU_-$ appears for $\eta = \eta_-$
and $\cQ_2$ comes back to stability.
The difference is that when $\eta$ increases up to $\eta = \eta_1$ both
equilibria $\cU_+$ and~$\cU_-$ disappear on~$\cQ_1$.
For $\eta > \eta_1$ the only remaining equilibria are the singular
ones, both stable.
Note that the bifurcation sequence resembles the passage
through resonance.
The possible configurations on the reduced phase space
section~\eqref{reduced phase section} are shown in
Fig.~\ref{Fig: X2ab} for increasing values of~$\eta$.

\item{Case \blue{III}. $0 < \alpha + 2 \lambda < \alpha$,
i.e.\ $\lambda < 0$}.
In this case the threshold values \eqref{instability y}
and~\eqref{instability x} are still all positive, however now
$\eta_1 < \eta_+ < \eta_-$ and the parabola~\eqref{parabola}
is convex, since $\lambda < 0$.
Therefore we see first the appearance of both equilibria
$\cU_-$ and~$\cU_+$ from the singular equilibrium at the origin.
Since~\eqref{parabola} is convex, the equilibrium~$\cU_-$ is stable
and $\cU_+$ is unstable now.
Such equilibria disappear then on~$\cQ_2$.
The first equilibrium to disappear is the one at~$\cU_+$, for
$\eta = \eta_+$, while the equilbrium~$\cQ_2$ becomes unstable.
At $\eta = \eta_-$ also the the equilibrium ~$\cU_-$ disappears and the
equilbrium~$\cQ_2$ turns back to stability. 
Also here the bifurcation sequence resembles the passage
through resonance.

\item{Case \blue{IV}. $\alpha + 2 \lambda < 0 < \alpha$.}
In this case the only acceptable threshold value is~$\eta_1$.
This implies that bifurcations can occur only from the equilibrium
at the origin.
At $\eta = \eta_1$ both equilibria $\cU_-$ and $\cU_+$
bifurcate off from the origin, the equilibrium~$\cU_+$ on
the lower arc is unstable and $\cU_-$ is stable.
No bifurcation occurs from the equilibrium at~$\cQ_2$, which is
always stable.
As in Case~\blue{I}, increasing $\eta$ beyond~$\eta_1$ merely
increases the size of~$\cP^{\eta}$, but does not change the
configuration of equilibria.

\item{Case \blue{V}. $\alpha < 0$ and $\alpha + 2 \lambda < 0$.}
All the critical values are not acceptable.
Therefore the only equilibria are $\cQ_1$ and $\cQ_2$, both stable.
In Fig.~\ref{fig: daDiagram} this case corresponds to the (empty)
left quadrants of the remaining two lower bifurcation diagrams.
\end{description}

\noindent
The regions on the $(\lambda, \alpha)$--plane corresponding to
the sequences~\blue{I}--\blue{V} are displayed in
Fig.~\ref{fig: daDiagram}, also for $\alpha + 2 \lambda < 0$. 
In this case, the difference $\eta_+ - \eta_-$ 
changes its sign and the parabola reverses its concavity. 
As a consequence the equilibria $\cU_+$ and~$\cU_-$ exchange
their stability and exchange themselves in the bifurcation sequence.
Moreover, all the inequalities on $\alpha, \delta, \lambda$ must be
inverted.
For example Case~\red{III} occurs now for
$\alpha < \alpha + 2 \lambda < 0$ and $\delta > 0$, with $\cU_+$
and~$\cU_-$ exchanging their role in the bifurcation sequence.
For $\delta > 0$ the cases shift to the red labeling in
Fig.~\ref{fig: daDiagram}.

For $\delta = 0$ the thresholds satisfy
$\eta_+ = \eta_- = \eta_1 = 0$ as all bifurcation lines originate
from the origin; recall that $\alpha \neq 0$ and
$\alpha + 2 \lambda \neq 0$ (next to $\lambda \neq 0$).
In Cases \blue{I} \&~\blue{IV} ($\alpha$ and $\alpha + 2 \lambda$
do not have the same sign) we also have for $\delta = 0$ the
configuration of equilibria $\cU_+$ and~$\cU_-$ next to $\cQ_1$
and~$\cQ_2$ as in Fig.~\ref{fig: daDiagram} (upper-left and
lower-right); otherwise ($\alpha$ and $\alpha + 2 \lambda$ have the
same sign) the situation is that of Case~\blue{V} except that the
critical values are all zero, i.e.\ at $\eta = 0$ the extreme of
the parabola~\eqref{reduced phase section} passes through~$\cQ_1$.

\subsection{Bifurcation mapping}
\label{sec:bifurcationmapping}

We have seen in the previous section that the bifurcation sequences
are determined by the detuning~$\delta$ and the coefficients
$\alpha, \lambda$.
The coupling constant $\mu$ may take any value, but the degenerate
case $\mu = 0$ is not included in the general approach; we could
easily scale $\mu = 1$ and conclude $\mu > 0$, but keep $\mu$ in the
formulas to allow for fast conclusions concerning reversible systems
with $\mu < 0$.
By the results of the previous section, the additional coefficients
$\beta_i$, $\gamma_j$, do not modify the qualitative picture.

\begin{figure}[htb]
\begin{center}
\begin{picture}(156,156)
   \put(10,0){\includegraphics[height=5cm,keepaspectratio]{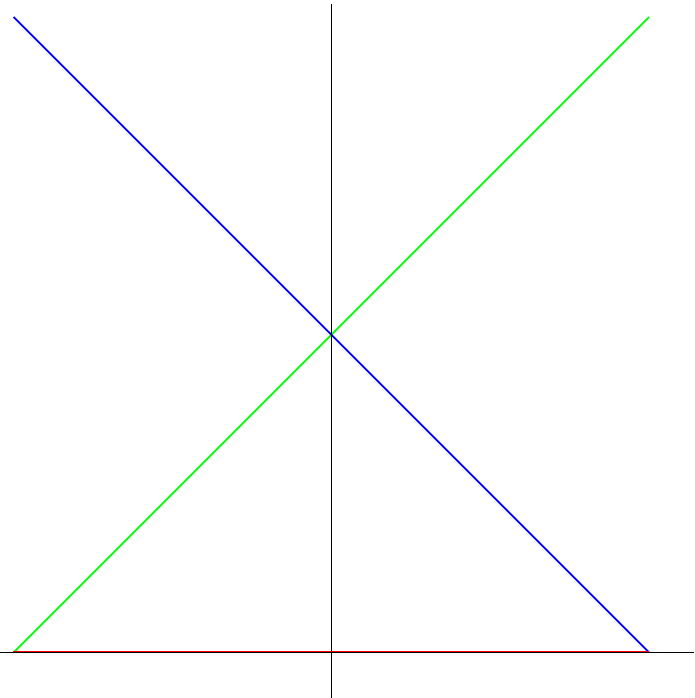}}
   \put(10,0){\includegraphics[height=5cm,keepaspectratio]{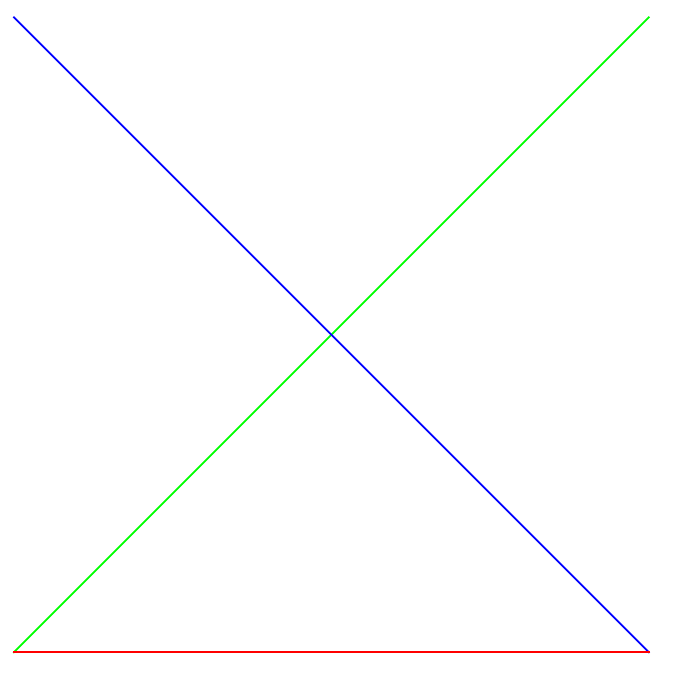}}
   \put(157,8){\footnotesize $C$}
   \put(74,147){\footnotesize $A$}
   \put(11,8){\tiny $|$}
   \put(5,-1){\footnotesize $-1$}
   \put(140.5,8){\tiny $|$}
   \put(139,-1){\footnotesize $1$}
   \put(73.5,71.3){\small $-$}
   \put(86,71){\footnotesize $1$}
\end{picture}
\end{center}
\caption{\small{The bifurcation plot in the $(C, A)$--plane,
see~\eqref{APCP}.
}}
\label{bplot}
\end{figure}
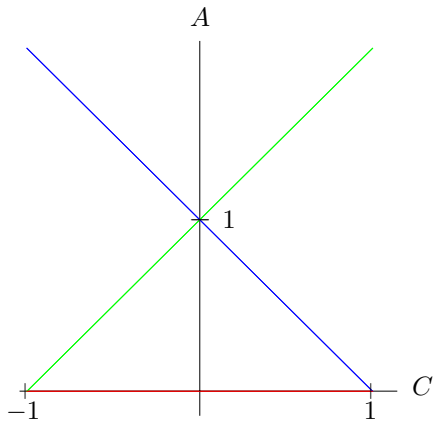

To give a more abstract view we can introduce --- in analogy
to what is done in~\cite{PM14} --- the parameters
\bs
\label{APCP}
\begin{align}
   A & \;\; := \;\; - \frac{2 \delta + \alpha \eta}{2 \lambda \eta}
\label{AP}\\
   C & \;\; := \;\; - \frac{\mu \delta}{2 \lambda (\alpha + 2 \lambda)}
   \enspace .
\label{CP}
\end{align}
\es
The first, the {\sl asymmetry} parameter, measures how far is the
system from the resonance manifold.
The second, the {\sl coupling} parameter, is a measure of the strength
of resonant coupling.
The Jacobi-determinant
\bd
\det D_{\delta, \eta} \left(
\begin{array}{c}
   C \\ A
\end{array} \right) \;\; = \;\; \det \left(
\begin{array}{cc}
   \frac{- \mu}{2 \lambda (\alpha + 2 \lambda)} & 0 \\
   \frac{-1}{\lambda \eta} & \frac{\delta}{\lambda \eta^2} 
\end{array}
\right) \;\; = \;\; \frac{C}{\lambda \eta^2}
\ed
of the bifurcation mapping warns us to be careful where $\delta$
and/or~$\eta$ vanish.
To capture the main qualitative features of the system we assume that
the non-vanishing parameters are $\delta, \alpha, \lambda, \mu$ and
put $\beta_i \equiv \gamma_j \equiv 0$.
The bifurcation thresholds then are
\ba
\eta_1 &=& - \frac{2 \delta }{\alpha}
\label{IYS}\\
\eta_{\pm} &=&
\frac{- 2 \delta }{\alpha + 2 \lambda } \; \mp \;
\frac{2 \mu \delta^2}{(\alpha + 2 \lambda)^3} \enspace .
\label{IXS}
\ea
We see that the asymmetry parameter vanishes at the critical value~\eqref{IYS},
\be
A(\eta_1) \;\; = \;\; 0
\label{A0}
\ee
and that, to first order in~$\delta$,
\be
A(\eta_{\pm}) \;\; = \;\; 1 \; \mp \;
\frac{\mu \delta}{2 \lambda (\alpha + 2 \lambda)}
\;\; = \;\; 1 \; \pm \; C
\enspace .
\label{Apm}
\ee
Then, we see that we can plot the straight lines \eqref{A0}
and~\eqref{Apm}, in the interval $-1\le C \le 1$, to get the
whole picture (see Fig.~\ref{bplot}).
We excluded the possibility of cases with $|C| > 1$, which is
equivalent to say that, at first order, $\eta_1$ cannot stay
between $\eta_+$ and~$\eta_-$.
For sufficiently small~$\delta$ this is ruled out by
the assumption $\lambda \neq 0$ that also ensures that $A$
and~$C$ are well defined.

In this plot, a vertical straight line represents a given system at
varying the distinguished parameter~$\eta$.
Therefore, we can recapitulate the bifurcation scenario in the light
of the plot in Fig.~\ref{bplot}.
Let us recall the five cases enumerated in the previous subsection.

\begin{description}
\item{Case \blue{I}. $\alpha < 0 < \alpha + 2 \lambda$,
$C > 0$ (right half-plane in Fig.~\ref{bplot}}).
Since the critical value $\eta_1$ is not acceptable, the red
horizontal line disappears from the plot.
Considering the parameter~\eqref{AP}, a sequence with growing~$\eta$
goes from top to bottom. 
Bifurcations occur when $\eta$ passes first through $\eta_{+}$
(green line), then through $\eta_-$ (blue line).
\item{Case \blue{II}. $0 < \alpha < \alpha + 2 \lambda$, $C > 0$.} 
All critical values are acceptable now and the sequence with
growing~$\eta$ still goes from top to bottom: 
Since $\eta_+ < \eta_- < \eta_1$, the sequence is given by
the green-blue-red passings.
\item{Case \blue{III}. $0 < \alpha + 2 \lambda < \alpha$,
$C < 0$ (left half-plane in Fig.~\ref{bplot}}).
All critical values are still acceptable but now
$\eta_1 < \eta_+ < \eta_-$ and the sequence with growing~$\eta$
goes from bottom to top: the sequence is now given by the
red-green-blue passings.
\item{Case \blue{IV}. $\alpha + 2 \lambda < 0 < \alpha$, $C > 0$.} 
Only $\eta_1$ is acceptable, whereas $\eta_{\pm}$ are both  not acceptable:
the only line present in the plot is the red one.
\item{Case \blue{V}. $\alpha < 0$ and $\alpha + 2 \lambda < 0$.} 
None of the thresholds  is acceptable and the plot is empty,
no bifurcations occur.
\end{description}

\begin{figure}[htb]
\center
\includegraphics[height=4.3cm,keepaspectratio]{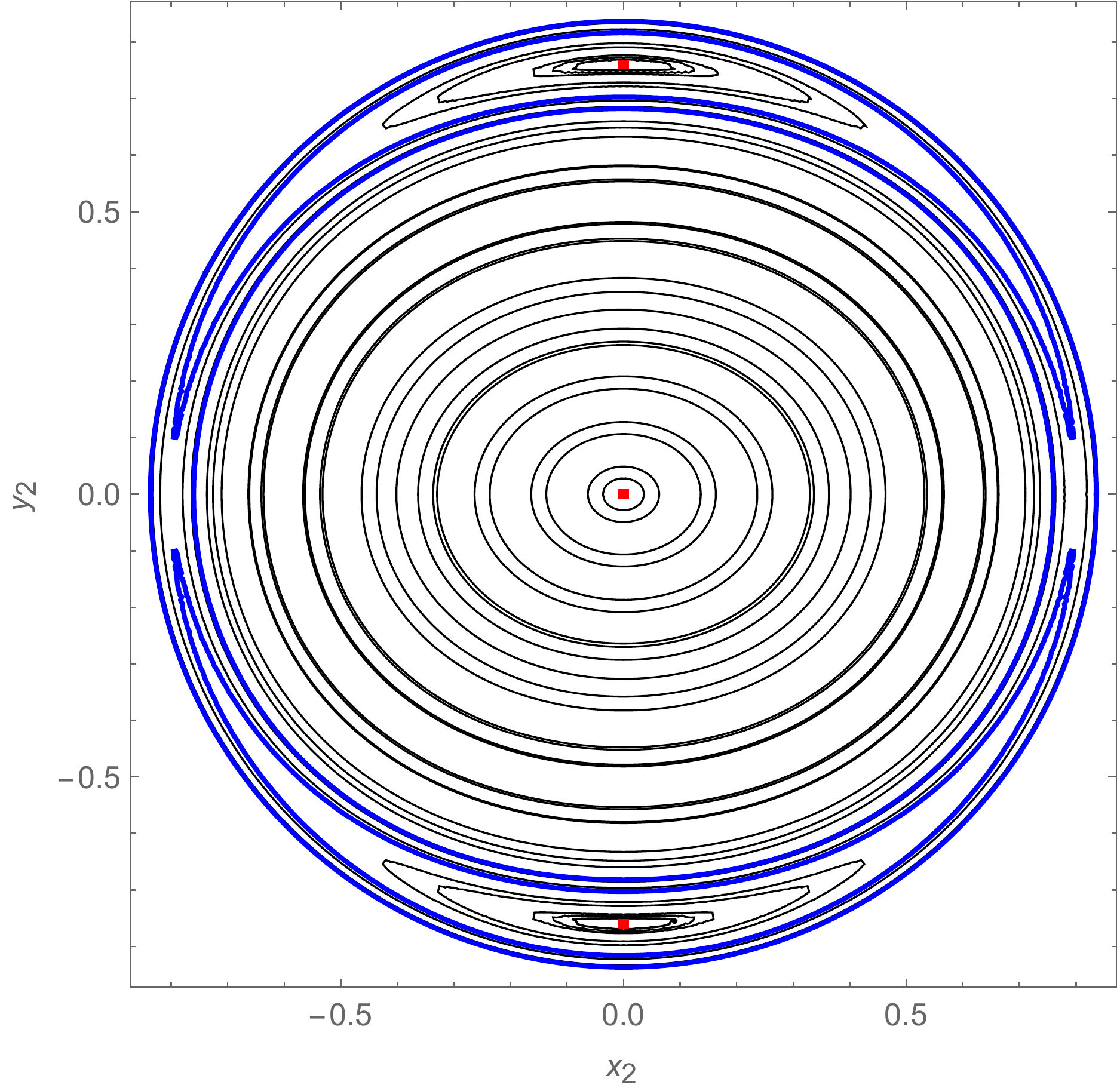}
\quad
\includegraphics[height=4.3cm,keepaspectratio]{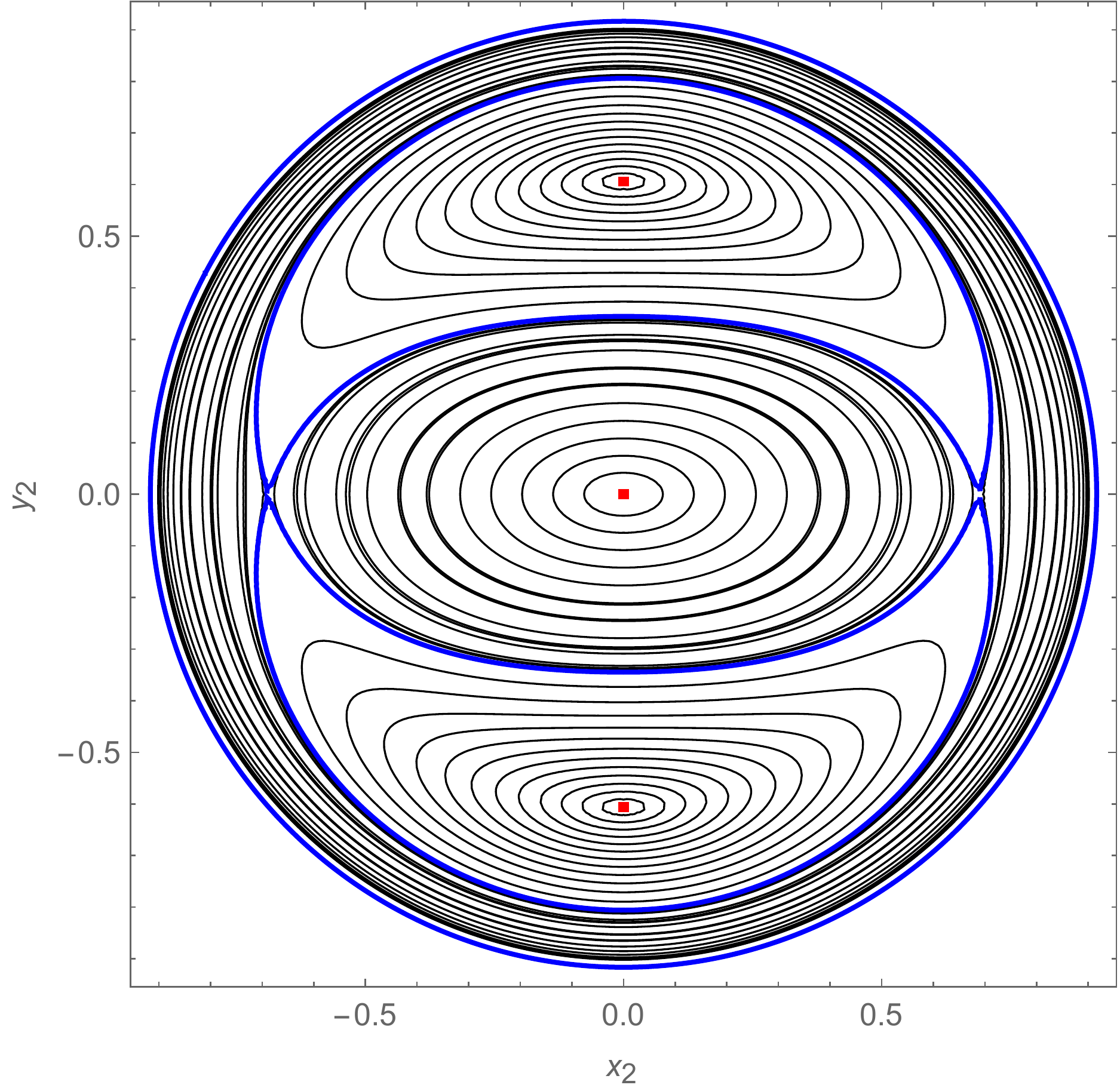}
\quad
\includegraphics[height=4.3cm,keepaspectratio]{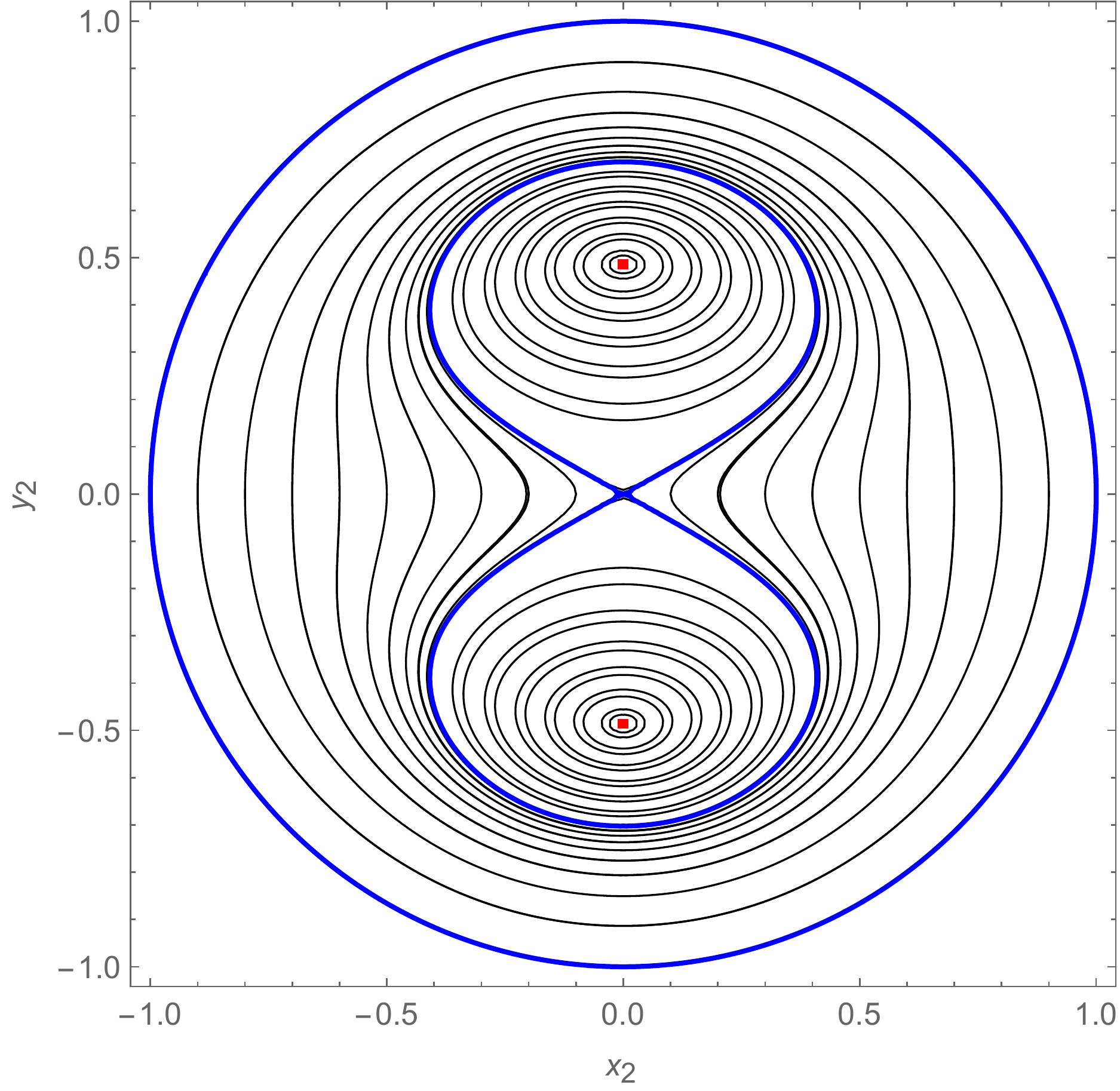}
\vskip0.3cm
\includegraphics[height=4.3cm,keepaspectratio]{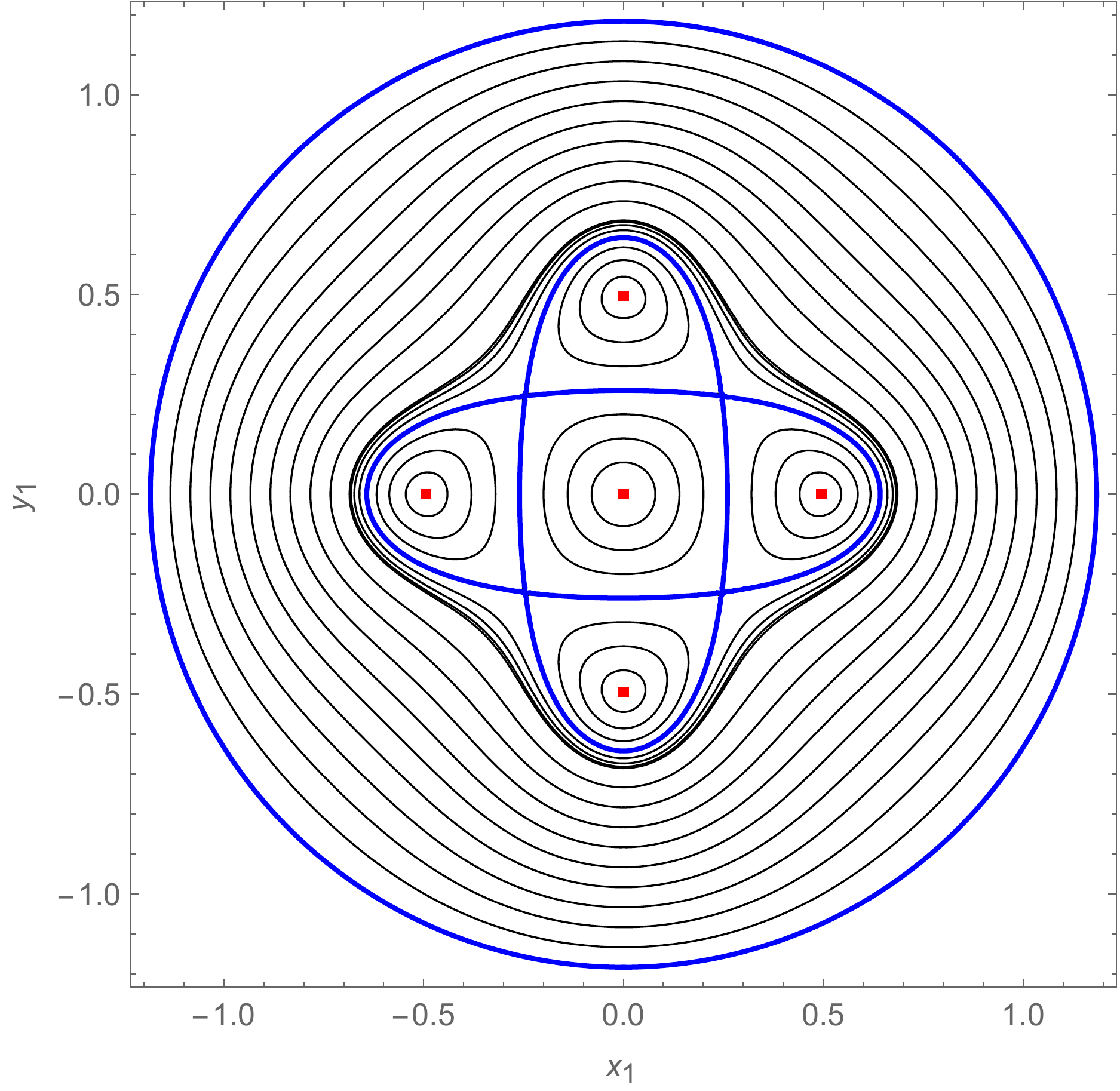}
\quad
\includegraphics[height=4.3cm,keepaspectratio]{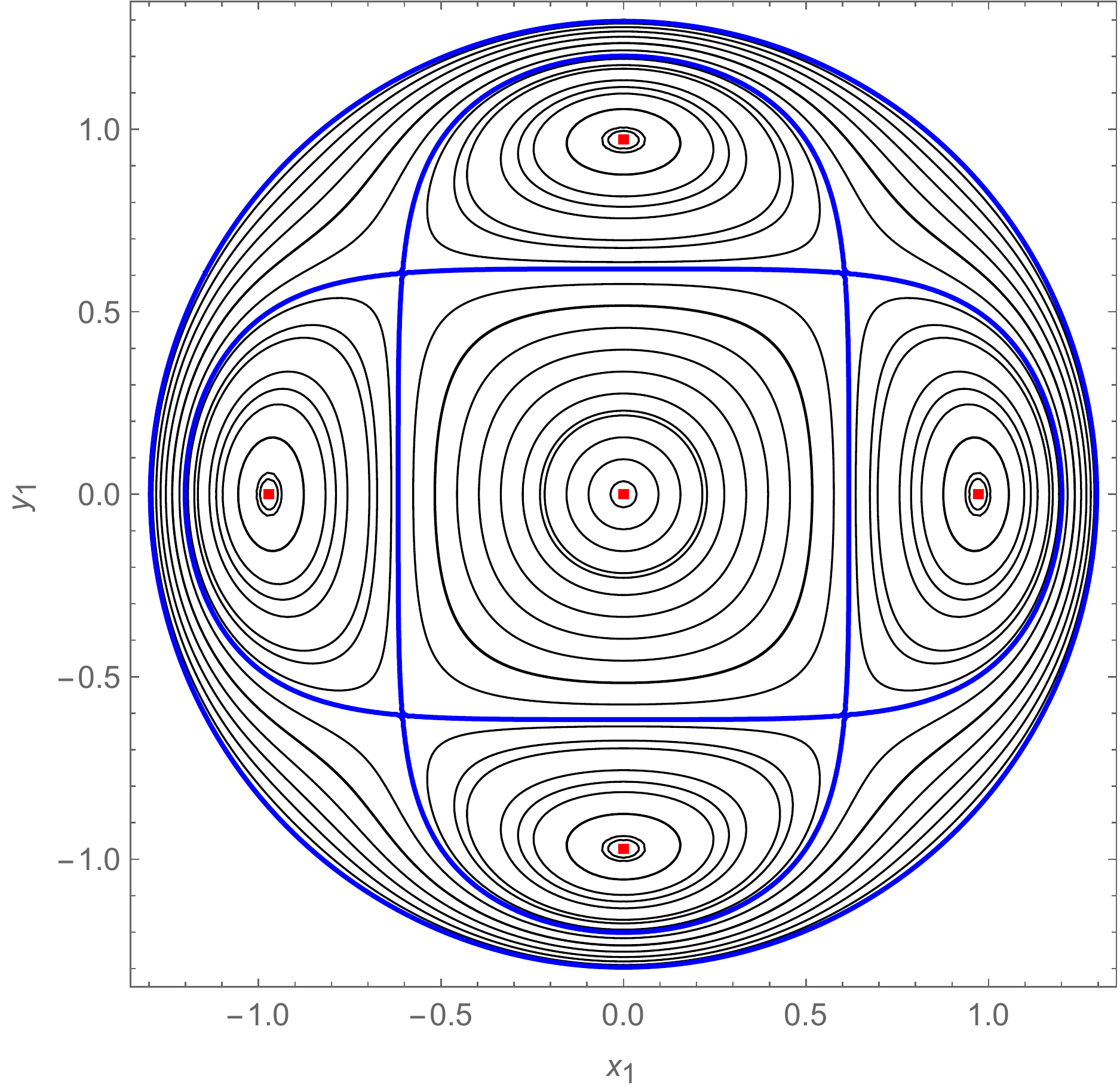}
\quad
\includegraphics[height=4.3cm,keepaspectratio]{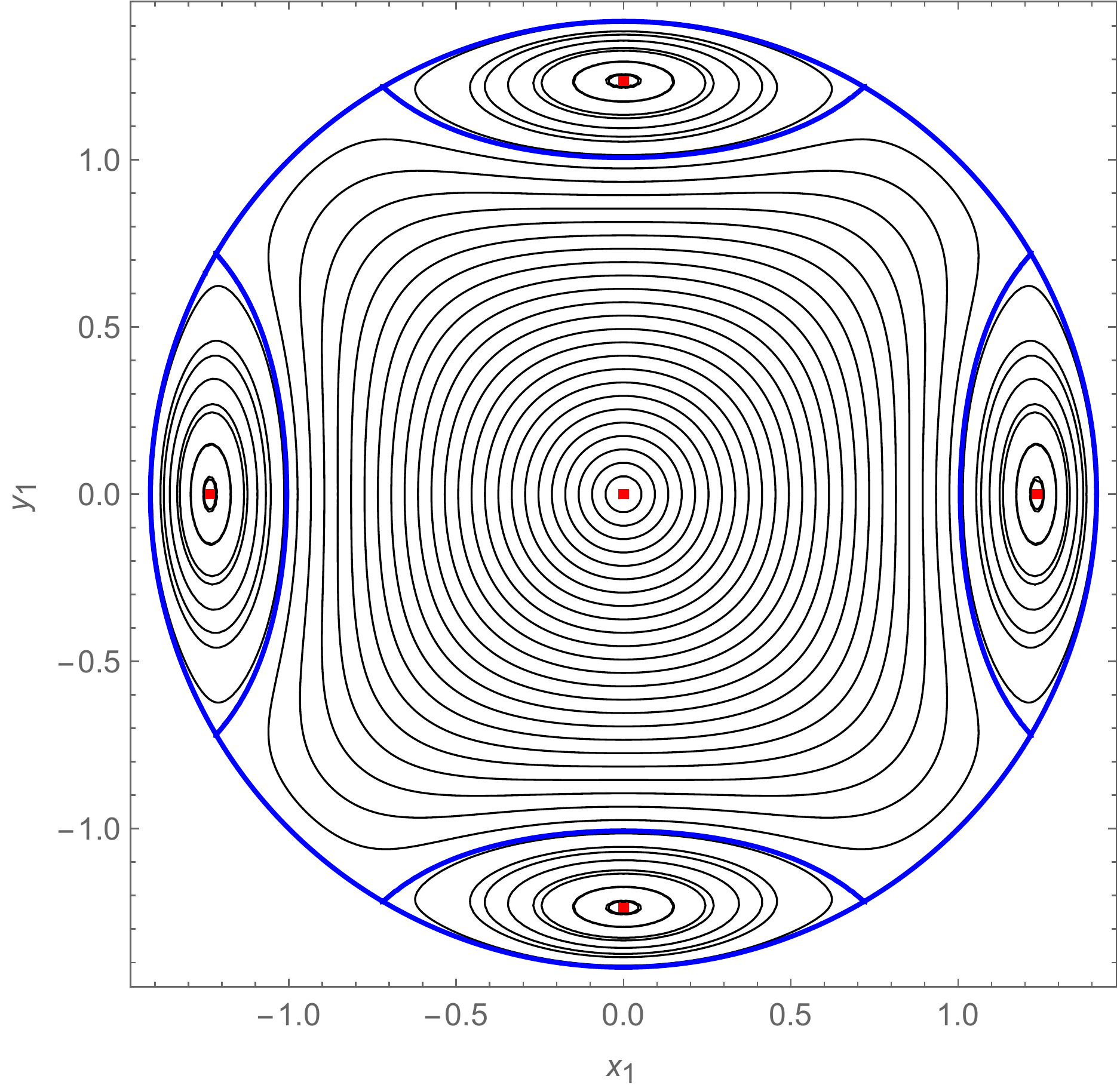}
\caption{\small{
Sequence of $x_2,y_2$ surfaces of section (above) and
corresponding $x_1,y_1$ surfaces of section (below) in Case~\blue{III}.
}\label{pss} }
\end{figure}

\noindent
In Fig.~\ref{pss}, a typical sequence of surfaces of section
corresponding to Case~\blue{III} is shown  to give an impression
on how the abstract Fig.~\ref{bplot} translates to the concrete
dynamics.
The surfaces of section of the other cases are different, but the
way how they relate to the corresponding part of Fig.~\ref{bplot}
is similar.

\section{Bifurcations in the original system}
\label{sec:bifurcations}

If a normalization is carried far enough to obtain only isolated
equilibria (after symmetry reduction), we know the essential
characteristics of the system.
Including higher orders may shift the positions of the equilibria,
but does not alter their number or stability.
Therefore the isolated fixed points of~\eqref{general normal form}
correspond to periodic orbits for the original system.
The results obtained can be trusted up to low energies,
in a neighbourhood of the central equilibrium and not too far from
the resonance at hand. 

To deduce the periodic orbits of the system from the equilibria
of~\eqref{general normal form}, we introduce
action angle variables
\be
\label{aa variables}
z_j \;\; = \;\; \sqrt{2 \tau_j} \, \E^{\I \varphi_j}
\enspace , \quad j = 1, 2 \enspace .
\ee
The singular equilibria correspond to the normal modes of the
system.
Namely, $\cQ_1$ corresponds to $\tau_1 = 0$ and
$\tau_2 = \frac{1}{2} \eta$, i.e.\ to the orbit along the
$x_2$--axis, in the following also referred to as short axial orbit.
Similarly, $\cQ_2$ gives the orbit along the $x_1$--axis, also
referred to as long axial orbit, determined by $\tau_1 = \eta$ and
$\tau_2 = 0$. 

\begin{remark}
To be consistent with our previous papers, so that the
conditions \eqref{banana orbits} and~\eqref{antibanana orbits} 
for banana and anti-banana are the same as in for
example~\cite{MP2013b}, one has to exchange sine and cosine
in the following.
\end{remark}

\noindent
The regular equilibria correspond to periodic orbits in general
position.
The equilibrium $\cU_+$ has co-ordinates $(u, v, w)$ such that
$0 < u < \eta$, $v < 0$ and $w = 0$.
From~\eqref{Z coordinates} we see that it must then be
$v = - \sigma_2^2$ and $\sigma_1 = 0$.
By expressing~\eqref{tau generators} in terms
of~\eqref{aa variables} we get the condition
\bd
\sigma_1 \;\; = \;\;
\tau_1 \sqrt{2 \tau_2}\, \cos(2 \varphi_1 - \varphi_2)
\;\; = \;\; 0
\enspace , \quad \tau_1 , 2 \tau_2 \in \opin{0}{\eta} \enspace .
\ed
This implies
\be
\label{banana orbits}
2 \varphi_1 \; - \; \varphi_2 \;\; \in \;\;
\left\{ \frac{\pi}{2}, \frac{3\pi}{2} \right\}
\enspace .
\ee
Similarly, we recognize that the equilibrium $\cU_-$ corresponds
to the condition 
\bd
\sigma_2 \;\; = \;\;
\tau_1 \sqrt{2 \tau_2}\, \sin(2 \varphi_1 - \varphi_2)
\;\; = \;\; 0
\enspace , \quad \tau_1 , 2 \tau_2 \in \opin{0}{\eta}
\ed
that gives
\be
\label{antibanana orbits}
2 \varphi_1 \; - \; \varphi_2 \;\; \in \;\;
\left\{ 0, \pi \right\}
\enspace .
\ee
Orbits satisfying \eqref{banana orbits} and~\eqref{antibanana orbits}
are called, because of their shape in the $(x_1, x_2)$--plane,
banana orbits and figure-eight or anti-banana orbits,
respectively~\cite{MS}.
We found in section~\ref{sec:singularequilibriaandtheirstability}
the critical values \eqref{instability y} and~\eqref{instability x}
that determine the bifurcations of the reduced system.
However, $\eta$ is not a constant for the original system;
nevertheless we can use \eqref{instability y}
and~\eqref{instability x} to find threshold values for the
bifurcations in terms of the (generalized) energy~$E$. 
On the long axial orbit ($\tau_1 = \eta$, $\tau_2 = 0$),
the normal form~\eqref{general normal form}
reads as
\be
\label{normal form on long axis}
K \;\; = \;\; \eta \; + \;
\varepsilon^2 (2 \delta + \alpha_1 \eta) \eta \; + \;
\varepsilon^4 (\rho_1 \delta + \alpha_6 \eta) \eta^2
\enspace .
\ee 
Here and in the following, since we refer to the original system,
we express the formulas in terms of the coefficients  of the
original normal form \eqref{general normal form}.
By the scaling of time~\eqref{time scaling} we have
\be
\label{energy scaling}
\frac{\omega_2}{2} K \; + \; O (\varepsilon^6)
\;\; = \;\; H \;\; = \;\; E
\ee
and combining \eqref{normal form on long axis}
and~\eqref{energy scaling} we can express
the (generalized) energy in terms of~$\eta$ as
\be
\label{energy on long axis}
E \;\; = \;\; \frac{\omega_2}2 \left[ \eta \, + \,
\varepsilon^2 \left( 2 \delta + \alpha_1 \eta \right) \eta
\, + \, \varepsilon^4 ( \rho_1 \delta + \alpha_6 \eta) \eta^2
\right] \; + \; O (\varepsilon^6) \enspace .
\ee
Substituting  \eqref{instability x} into~\eqref{energy on long axis},
we find the critical energy threshold values that correspond to the
bifurcations off from the long axial orbit, given to second order
in~$\delta$ by
\be
\label{E instability x}
E_{\pm} \;\; := \;\;
- \frac{2 \omega_2 \delta}{4 \alpha_1 - \alpha_3} \; + \;
\frac{ 4 \omega_2 \delta^2}{(4 \alpha_1 - \alpha_3)^3}
\, \left[ (4 \alpha_1 - \alpha_3)
(2 \beta_1 + \beta_2 - 2 \alpha_1 + \alpha_3)
\, - \, 2 (\gamma \pm \mu) \right]
\ee
for banana~(upper signs) and anti-banana~(lower signs)
orbits, respectively.
Recall $\beta_1 = \rho_1 + \frac{1}{4} \rho_2 - \frac{1}{2} \rho_3$
and $\beta_2 = \frac{1}{2} \rho_3 - \frac{1}{2} \rho_2$
from~\eqref{new coefficients lab} and from~\eqref{instability x} that
$\gamma = 6 \gamma_1 + 4 \gamma_2 + 2 \gamma_3 = 6 \alpha_6 - \alpha_4$.
We inverted the detuning scaling~\eqref{detuning scaling}, so
that \eqref{E instability x} and~\eqref{E instability y} are
expressed in terms of the original detuning parameter.
Similarly, for the bifurcation off from the short axial
orbit ($\tau_1 = 0$, $\tau_2 = \frac{1}{2} \eta$) we use~\eqref{instability y} 
and find to second order in~$\delta$ that
\be
\label{E instability y}
E_{1} \;\; = \;\; - \frac{2 \omega_2 \delta}{\alpha_3 - \alpha_2}
\; + \; \frac{2 \omega_2 \delta^2}{(\alpha_3 - \alpha_2)^3}
\,  \left[ (\alpha_3 - \alpha_2)
(2\beta_2 + \alpha_2) \, - \, 8 \gamma_3 \right]
\ee
where $\beta_2 = \frac{1}{2} \rho_3 - \frac{1}{2} \rho_2$ and
$\gamma_3 = \frac{1}{4} \alpha_5 - \frac{3}{8} \alpha_7$.
However, above a certain threshold one should not expect that the
formal series developed by the normalization procedure stays close
for a very long time to the solutions of the original problem.
Since we pushed the normalization up to including terms of $6$th
order in the phase space variables~$(x, y)$, we can trust such
quantitative predictions on the bifurcation and stability of the
periodic orbits  up to the second order in the detuning parameter
(since, we recall, this is assumed to be a second order term).
We can summarize these results as follows.

\begin{theorem}
\label{thm: instability normal modes}
Let us consider the dynamical systems defined by~$H$,
cf.~\eqref{Hamiltonian series} and its normal
form~\eqref{general normal form} with respect to the oscillator
symmetry~\eqref{oscillator symmetry}.
Assume the co-ordinate system  to be rotated so that $\mu > 0$
and $\nu = 0$ in~\eqref{general normal form}.
In a neighbourhood of the central equilibrium and for sufficiently
small values of the detuning parameter~$\delta$,
\begin{itemize}
\item[i)]
at the stability/instability transition of a normal mode, periodic
orbits in general position bifurcate.
In particular, at each transition of the long axial orbit, a pair
of periodic orbits bifurcate (a pair of banana or a pair of
anti-banana orbits).
At the instability  of the short axial orbit, two pairs of periodic
orbits (a pair of banana and a pair of anti-banana orbits) bifurcate
concurrently;
\item[ii)]
up to second order in the detuning, the instability/stability
transition of the normal modes occur at the critical energies
\eqref{E instability x} and~\eqref{E instability y} for the long
and short normal mode, respectively.
\end{itemize}
\end{theorem}

\noindent
The coefficients $\alpha$ and~$\lambda$ determine the possible
bifurcation sequences according to the previous section.
Recalling that $2 (\alpha + 2 \lambda) = 4 \alpha_1 - \alpha_3$ 
and $2 \alpha = \alpha_3 - \alpha_2$, the analysis that resulted
into Cases \blue{I}--\blue{V} can be rewritten in terms of the
periodic orbits of the original system.

\begin{theorem}
\label{thm: bifurcation sequences}
Under the conditions of theorem~\ref{thm: instability normal modes}
the possible bifurcation sequences are determined by the coefficients
$\alpha_1$, $\alpha_2$, $\alpha_3$ in the normal
form~\eqref{general normal form}.
For $\alpha_3 \neq \alpha_2$,
$\alpha_3 \neq 2 \alpha_1 + \frac{1}{2} \alpha_2$,
$\alpha_3 \neq 4 \alpha_1$ and $\delta < 0$ we have the following cases.
\begin{description}
\item{Case \blue{I}. $\alpha_3 - \alpha_2 < 0 < 4 \alpha_1 - \alpha_3$.}
The short axial orbit is always stable.
The long axial orbit changes its stability twice.
At first it suffers a transition to instability at the critical
energy $E = E_+$ and a pair of stable banana orbits appears.
At $E = E_-$ a pair of anti-banana orbits appears, while the
long axial orbit comes back to stability.
\item{Case \blue{II}. $0 < \alpha_3 - \alpha_2 < 4 \alpha_1 - \alpha_3$.}
While the (generalized) energy passes through the critical values
$E = E_+$ and subsequently $E = E_-$, the bifurcation sequence
follows the previous case.
However a further bifurcation occurs at $E = E_1$, when both pairs
of periodic orbits in general position disappear on the short axial
orbit.
\item{Case \blue{III}. $0 < 4 \alpha_1 - \alpha_3 < \alpha_3 - \alpha_2$.}
At $E = E_1$ a pair of banana and a pair of anti-banana orbits
bifurcate off from the short axial orbit.
The banana orbits are unstable and the anti-banana orbits are stable.
At $E = E_+$ banana orbits disappear on the long axial orbits,
which becomes unstable.
At $E = E_-$ anti-banana orbits disappear as well, and the long axial
orbit turns back to stability.
\item{Case \blue{IV}. $4 \alpha_1 - \alpha_3 < 0 < \alpha_3 - \alpha_2$.}
At $E = E_1$ a bifurcation occurs from the short axial orbit, and
the two pairs of periodic orbits in general position appear.
Banana orbits are unstable and anti-banana orbits are stable.
The long axial orbit is always stable.
\item{ Case \blue{V}: $4 \alpha_1 < \alpha_3 < \alpha_2$.}
The only periodic orbits are the normal modes, both stable.
\end{description}
The undetuned system for $\delta = 0$ behaves as in Cases \blue{I}
or~\blue{IV} (after the bifurcations) if $\alpha_3 - \alpha_2$ and
$4 \alpha_1 - \alpha_3$ have opposite signs and otherwise as in
Case~\blue{V}.
\end{theorem}

\noindent
Assuming the co-ordinate system to be such that $\nu$ vanishes in
the normal form~\eqref{general normal form} is not needed for
qualitative predictions, but only for quantitative ones.
And even here the necessary rotation can simply be turned back.
In fact, the presence of~$\nu$ would not change the possible
bifurcation scenario of the system, but would affect the value of
the energy thresholds \eqref{E instability x}
and~\eqref{E instability y} --- replacing $\mu$ by
$\sqrt{\mu^2 + \nu^2}$. 

\begin{remark}
\label{remark:bif_order}
Let us note one more time that for definiteness we assumed $\mu$
positive and $\delta$ negative. 
Taking $\delta > 0$ yields the red cases~\red{I}--\red{V}; as a
demonstration see the example in
section~\ref{galacticdynamicsunderpowerlawpotentials}.
Since the difference in the bifurcation
thresholds~\eqref{E instability x} is proportional to
$\mu (4 \alpha_1 - \alpha_3)^{-3}$, a change in the
sign of~$\mu$ would affect only the bifurcation order of banana
and anti-banana orbits, that consequently would also exchange
their stability properties, as would a change in the sign of
$4 \alpha_1 - \alpha_3$.
\end{remark}

\section{Galactic dynamics under power law potentials}
\label{sec:example}
\label{galacticdynamicsunderpowerlawpotentials}

To demonstrate our results with an example, let us consider
the family of potentials
\ba
\label{pota}
V (x_1, x_2; q, p) \;\; = \;\;
\frac{1}{p} \left( 1 + x_1^2 + \frac{x_2^2}{q^2} \right)^{p/2}
\enspace , \quad  \mbox{$0 < p < 2$, $\frac{1}{4} < q \leq 1$}.
\ea
This gravitational potential is generated by a simple but realistic
matter distribution~\cite{scu95, PL96, TT97, scu99, BBP2}.
Its astrophysical relevance~\cite{bin81, BT08} is based
on the ability to describe in a simple way the gross features of
elliptical galaxies embedded in a dark matter halo.
Here the lower limit $p \to 0$ corresponds to the logarithmic
potential, while the upper limit $p \to 2$ is dictated by the
assumption of a potential generated by a positive mass distribution.

The flattening is~$1/q$ and we slightly extend its range from the range
$\frac{1}{2} < q \leq 1$ used in~\cite{MP2013b} that is typically associated
with elliptical galaxies~\cite{MS} as it is precisely at
$q = \frac{1}{2}$ that the $2{:}4$~resonance occurs.
Lower positive values of $q$ can in principle be considered but
correspond to an unphysical density distribution.
The truncated series expansion~\eqref{Hamiltonian series} is
``prepared'' for normalization by setting
\be
\label{qdelta}
q \;\; = \;\; \frac{\omega_1}{\omega_2} \;\; = \;\; \frac{1}{2} \; + \; \delta
\enspace ,
\ee
the canonical variables and time are rescaled according to~\eqref{scaling}
and we expanded in series of the detuning according to~\eqref{qdelta}.
The coefficients of the normal form~\eqref{general normal form} read as
\bs
\label{app1: normal form coefficients}
\begin{align}
&
\alpha_1 =\frac32 B_1 ,\;\;
\alpha_2 =6 B_1 ,\;\;
\alpha_3 =4 B_1,\;\;
\rho_1= 3 B_1,\;\;\rho_2=-12 B_1,\;\;\rho_3=0
\label{app1: normal form coefficients ar}\\
&
\alpha_4 = - \frac{56}{3}B_1^2+ 9 B_2 ,\;\;
\alpha_5 = - \frac{2}{3}(46 B_1^2-27 B_2),\;\;
\label{app1: normal form coefficients a45}\\
&
\alpha_6 = \frac{17 B_1^2-10 B_2}{4},\;\;
\alpha_7 = -2 (17 B_1^2-10 B_2),\;\;
\label{app1: normal form coefficients a67}\\
&
\mu \; = \; 3 (2B_1^2-B_2) , \quad \nu \; = \; 0 \enspace ,
\label{app1: normal form coefficients mn}
\end{align}
\es
where 
\be
\label{Bcoeff}
B_1 \; = \; \frac{p-2}{8}
\quad \mbox{and} \quad
B_2 \; = \; \frac{(p-2)(p-4)}{48}
\enspace ,
\ee
compare with~\cite{MP2013b}.
As the potential is scalar, the ensuing system is reversible with respect to
\be
\label{time symmetry}
(x_1, x_2) \;\; \mapsto \;\; (x_1, -x_2)
\ee
which through reduction turns into
\bd
(u, v, w) \;\; \mapsto \;\; (u, v, -w)
\ed
and explains why $\nu = 0$.
By substituting \eqref{app1: normal form coefficients}
into~\eqref{new coefficients} we find
\be
\label{app1: lambda alpha}
\lambda \; = \; - \alpha \; = \; \frac{p-2}{8} \; < \; 0
\quad \mbox{and} \quad
\mu \; = \; \frac{1}{32}(p^2 - 4) \; < \; 0
\enspace .
\ee
Note that the $\pi$--rotation $(u, v, w) \mapsto (u, -v, -w)$ still
allows to achieve $\mu > 0$ in~\eqref{app1: lambda alpha}, if necessary.
According to theorem~\ref{thm: bifurcation sequences} and
remark~\ref{remark:bif_order}, the coefficients $\alpha_j$, $j=1,2,3$ 
determine the bifurcation sequences.
Since $4 \alpha_1 - \alpha_3 < 0 < \alpha_3 - \alpha_2$ and,
concentrating on $q > \frac{1}{2}$, the detuning $\delta$ is
positive, the bifurcation sequence follows Case~\red{I} of
theorem~\ref{thm: bifurcation sequences} (in which remember to
reverse all the inequalities, according to
remark~\ref{remark:bif_order}). 
As $\mu$ is negative, bifurcations occur always from the long normal
mode, with bananas appearing at lower energies than anti-bananas.
The critical values of the energy that determine the bifurcations
can be found by substituing \eqref{app1: normal form coefficients mn}
and \eqref{Bcoeff} into~\eqref{E instability x} and, expressed in
terms of the parameters of~\eqref{pota}, in agreement
with~\cite{MP2013b} read as
\bs
\label{critical hab}
\begin{align}
E_{+} & \;\; = \;\; \frac{16}{2-p} \left( q - \frac{1}{2} \right) \; + \;
\frac{8 (41 p - 10)}{3 (p-2)^2} \left( q - \frac{1}{2} \right)^2
\label{critical h2a}\\
E_{-} & \;\; = \;\; \frac{16}{2-p} \left( q - \frac{1}{2} \right) \; + \;
\frac{8 (53 p + 14)}{3 (p-2)^2} \left( q - \frac{1}{2} \right)^2
\label{critical h2b}
\end{align}
\es
for the bifurcation of banana and anti-banana orbits, respectively.
Numerical values of the thresholds when applied e.g.\ to the 
logarithmic potential (taking $p = 0$), are in good agreement
with the bifurcation values obtained from numerical
computations~\cite{MS}.

\begin{remark}
When modeling the dynamics in a rotating galaxy using the Hamiltonian
function
\begin{equation}
\label{rotatingpotential}
   H(x, y) \;\; = \;\;
   \frac{y_1^2 + y_2^2}{2} \; - \; \Omega (x_1 y_2 - x_2 y_1) \; + \;
   V (x_1, x_2; q, p)
   \enspace ,
\end{equation}
we generalize~\eqref{pota} which is the limit
of~\eqref{rotatingpotential} as $\Omega \to 0$.
Due to the rotation of the galaxy, the
Hamiltonian~\eqref{rotatingpotential} does not respect the
symmetries~\eqref{symmetries}. 
However, after diagonalization of the quadratic part, its 
series expansion still has the form~\eqref{Hamiltonian series}.
With the assumption that the angular velocity~$\Omega$  is a small
parameter and $\frac{1}{4} < q \leq 1$ as above,  the system can
again be studied as a perturbation of an oscillator close to a
$1{:}2$~resonance.
Since all the terms not respecting the symmetries~\eqref{symmetries} 
do not Poisson commute with the $1{:}2$ oscillator and odd order terms
are not present in the series expansion
of~\eqref{rotatingpotential}, a normalization of the
(truncated) diagonalized Hamiltonian then results in the normal form
of a $2{:}4$ (detuned) resonance, thus still of the
form~\eqref{general normal form}.
Compared with~\eqref{pota}, the presence of quartic terms of odd
order in the momenta in the diagonalized Hamiltonian produces
non-vanishing~$\nu$.
Modulo a rotation to eliminate~$\nu$,
theorems~\ref{thm: instability normal modes} and
\ref{thm: bifurcation sequences} can be applied.
The resulting families of periodic orbits would however correspond
to more ``fancy'' orbits for the original
system~\eqref{rotatingpotential}, once the diagonalizing
transformation is inverted.
We leave a deeper analysis of this problem to future work.
\end{remark}

\noindent
Several results of the theory developed above can be extended to
a $3$--dimensional model of the form
\begin{equation}
\label{3DHam}
   H(x, y) \;\; = \;\;
   \frac{y_1^2 + y_2^2 + y_3^2}{2} \; + \;
   V (x_1, x_2, x_3)
   \enspace ,
\end{equation}
in the cases in which the mirror symmetries~\eqref{symmetries} are
extended to the third axis when composing with the transformation law
\begin{equation}
\label{symmetry-3}
   (x_3 ,y_3) \;\; \mapsto \;\; (-x_3, -y_3) \enspace .
\end{equation}
Each symmetry plane of the potential generates an invariant subset
where the dynamics essentially reduce to those investigated above.
By introducing a further detuning parameter associated to the second
frequency ratio, bifurcation and stability of periodic orbit families
on the symmetry planes can be deduced. 
The validity of this approach is supported by analogous results
obtained with the $1{:}1{:}1$~resonance \cite{tdz, CFH99, FPY}. 
For a deeper understanding, $3$--dimensional normal forms of the
symmetric $1{:}1{:}2$, $1{:}2{:}2$ and $1{:}2{:}4$~resonances are
necessary~\cite{aa, aav, SVM07}, which usually provide the properties
of periodic orbits in general position.
Note that a normal form of the $1{:}2{:}2$~resonance is always
integrable, while already the cubic normal forms of the
$1{:}1{:}2$ and $1{:}2{:}4$~resonances are not integrable~\cite{chr}.
However, the discrete symmetries not only make the cubic terms vanish
but may furthermore enforce some of the non-trivial normal forms to
be integrable, see~\cite{HMV}.

\section{Conclusions}
\label{sec:conclusions}

We considered families of Hamiltonian systems in two degrees of
freedom with an equilibrium in $2{:}4$~resonance, a $1{:}2$~resonance
with additional discrete symmetry.
Under detuning, this typically leads to normal modes losing their
stability through period-doubling bifurcations.
This now concerns the long axial orbit, losing and regaining
stability through two period-doubling bifurcations.
In galactic dynamics one speaks of banana and anti-banana
orbits.
The short axial orbit turns out to be dynamically stable everywhere
except at a simultaneous bifurcation of banana and anti-banana orbits.

We excluded the case $\mu = 0$ from our considerations since it
would require further normalization.
Indeed, for $\mu = 0$ the normal form~\eqref{general normal form}
resembles~\eqref{reducedHamiltonianwepszero} and leads to a similar
degeneracy, which to break requires higher order terms that {\em do}
depend on $v$ (or on~$w$).
One may speculate that for such a $k{:}2k$~resonance, $k \geq 3$
also the conical singularity~$\cQ_2$ ``turns into'' a cusp and the
two successive period-doubling bifurcations of the long periodic
orbit occur simultaneously, as it happens to the two successive
period-doubling bifurcations of the short periodic orbit when
the $1{:}2$~resonance becomes the $2{:}4$~resonance.

\section*{Acknowledgements}
We thank the referees for the suggested improvements of the text. 
A.M. was supported by the Grant Agency of the Czech Republic,
project 17-11805S. G.P. acknowledges GNFM-INdAM and INFN for partial support.

\end{document}